\documentclass[11pt]{article}

\usepackage{latexsym,color,amsmath,amssymb,amsfonts,mathrsfs,euscript}
\usepackage{enumerate}
\usepackage{color}
\usepackage{graphicx}
\usepackage{theorem}
\usepackage{amstext}
\usepackage{amssymb}
\usepackage{url}

\newtheorem{theorem}{Theorem}[section]
\newtheorem{lemma}[theorem]{Lemma}
\newtheorem{corollary}[theorem]{Corollary}
\newtheorem{proposition}[theorem]{Proposition}
{\theorembodyfont{\rm} \newtheorem{defn}[theorem]{Definition}}

\newcommand{\Tree}{\EuScript{T}}
\newcounter{itemcounter}

\def\reals{\mathbb R}
\def\peps{(p,\varepsilon)}
\def\eps{\varepsilon}
\def\R{\cal{R}}
\def\r{\frak{r}}
\def\meas#1#2{\overline{#1}(#2)}
\def\ol#1{\overline{#1}}

\title{Output-Sensitive Tools for Range Searching in Higher Dimensions\thanks{%
Work by Micha Sharir was supported by NSF Grant CCF-08-30272, by Grants
2006/194 and 2012/229 from the U.S.-Israel Binational Science Foundation,
by Grants 338/09 and 892/13 from the Israel Science Fund, by the Israeli
Centers for Research Excellence (I-CORE) program (Center no. ~4/11), and
by the Hermann Minkowski--MINERVA Center for Geometry at Tel Aviv University.
The paper is part of the second author's M.Sc. dissertation, prepared under the supervision of the first author.
}}

\author{ Micha Sharir\thanks{%
School of Computer Science,
Tel Aviv University, Tel Aviv 69978, Israel, and
Courant Institute of Mathematical Sciences,
New York University, New York, NY 10012, USA;
\texttt{michas@tau.ac.il}.}
\and
Shai Zaban\thanks{%
School of Computer Science,
Tel Aviv University, Tel Aviv 69978, Israel;
\texttt{shaiza@bezeqint.net}.}}

\begin{document}
\maketitle

\begin{abstract}
Let $P$ be a set of $n$ points in $\reals^{d}$. A point $p \in P$ is $k$\emph{-shallow} if it lies in a halfspace which contains at most $k$ points of $P$
(including $p$). We show that if all points of $P$ are $k$-shallow, then $P$ can be partitioned into $\Theta(n/k)$ subsets, so that any hyperplane crosses
at most $O((n/k)^{1-1/(d-1)} \log^{2/(d-1)}(n/k))$ subsets. Given such a partition, we can apply the standard construction of a spanning tree with small
crossing number within each subset, to obtain a spanning tree for the point set $P$, with crossing number $O(n^{1-1/(d-1)}k^{1/d(d-1)}
\log^{2/(d-1)}(n/k))$. This allows us to extend the construction of Har-Peled and Sharir \cite{hs11} to three and higher dimensions, to obtain, for any set
of $n$ points in $\reals^{d}$ (without the shallowness assumption), a spanning tree $T$ with {\em small relative crossing number}. That is, any hyperplane
which contains $w \leq n/2$ points of $P$ on one side, crosses $O(n^{1-1/(d-1)}w^{1/d(d-1)} \log^{2/(d-1)}(n/w))$ edges of $T$. Using a similar mechanism,
we also obtain a data structure for halfspace range counting, which uses $O(n \log \log n)$ space (and somewhat higher preprocessing cost), and answers a
query in time $O(n^{1-1/(d-1)}k^{1/d(d-1)} (\log (n/k))^{O(1)})$, where $k$ is the output size.
\end{abstract}

\section{Introduction}

\subsection{Background}
One of the central themes in computational geometry is the design of efficient range searching algorithms. Typically, in such a problem one is given an
input set $P$ of $n$ points in $d$ dimensions, and the goal is to preprocess $P$ into a data structure, so that for any query {\em range} of some type (say,
a halfspace $H$), one can efficiently count, report, or test for emptiness the set $P \cap H$. A significant component of most of these algorithms involves
space decomposition techniques, which partition the input set into subsets with some useful structure. Apart from range searching, partitions are also used
in constructions of spanning trees with small crossing number, approximations, $\eps$-nets, and in several other computational geometry problems.

In a typical approach (see, e.g., \cite{mat92a}), given a point set $P$ in $\reals^{d}$, one partitions $P$ into subsets of approximately equal sizes, so
that each of them is contained within some region of constant description complexity (e.g., simplices). We remark that the subsets are pairwise disjoint,
while the containing regions might have nonempty intersections. We seek partitions with {\em small crossing number}, meaning that the maximum number of
enclosing regions crossed by the boundary of a range (typically, a hyperplane) is small.

In this paper, we consider a variant of the partition paradigm, referred to as a partition of a {\em shallow point set}, which is applied to a set $P$ of
{\em shallow} points in $\reals^{d}$, where a point $p \in P$ is $k$-shallow if there exists a hyperplane $h$ that contains at most $k$ points of $P$,
including $p$, on one of its sides. We will construct a partition whose crossing number, namely, the maximum number of its simplices that can be crossed by
a hyperplane, depends, in addition to $n$, also on the shallowness of $P$, i.e., on the maximum shallowness $k$ of its points.

As is typical in computational geometry, we consider the dimension $d$ as a fixed (small) constant, thus factors depending only on $d$ will be regarded as constants.


\paragraph{Cuttings.}
One of the major ingredients of our partitioning technique is {\em cuttings} of arrangements of hyperplanes. A cutting is a collection of (possibly
unbounded) $d$-dimensional closed cells with constant description complexity (e.g., simplices) with pairwise disjoint interiors, which cover the entire
$\reals^{d}$ (or some specified portion thereof). Let $H$ be a collection of $n$ hyperplanes in $\reals^{d}$ and let $\Xi$ be a cutting of the arrangement
$\mathcal{A}(H)$. For each simplex $\Delta \in \Xi$, let $H_{\Delta}$ denote the collection of hyperplanes intersecting the interior of $\Delta$. The
cutting $\Xi$ is called a $(1/r)$-{\em cutting} for $H$ if $|H_{\Delta}| \leq n/r$ for every simplex $\Delta \in \Xi$.

It will sometimes be convenient to work with weighted collections of hyperplanes, where such a collection is a pair ($H,w$), where $H$ is a collection of
hyperplanes, and $w:H \rightarrow \reals^{+}$ is a weight function on $H$. For each $L \subseteq H$, we write $w(L)$ for $\sum_{h \in L} w(h)$. The notions
introduced for unweighted collections of hyperplanes can usually be generalized for weighted collections in an obvious way. For example, a cutting $\Xi$ is
a $(1/r)$-cutting for ($H,w$) if for every simplex $\Delta \in \Xi$, the collection $H_{\Delta}$ has total weight at most $w(H)/r$.


\paragraph{Partitions: related work.} Our study extends to higher dimensions the planar construction of a partition of a set of shallow points, recently
developed by Har-Peled and Sharir \cite{hs11}. This technique partitions $P$ into $O(n/k)$ subsets, each containing $O(k)$ points and enclosed in a
(possible unbounded) triangle, so that the triangles are pairwise disjoint, and the crossing number of this partition (by any line) is only $O(\log(n/k))$.

Har-Peled and Sharir also constructed a set $P$ of $n$ $k$-shallow points in $\reals^{3}$ (in fact, $P$ is a set in convex position, so all its points are
$1$-shallow), so that the (maximum) crossing number of any partition of $P$ into subsets of size $\Theta(k)$ is $\Omega(\sqrt{n/k})$. See below for a
further discussion of this phenomenon, which was the starting point and the motivation for the study presented in this paper.

We next review the more standard partitioning techniques of Matou\v{s}ek \cite{mat92a,mat92b}. There $P$ is a set of $n$ points in $\reals^{d}$, $d \geq 2$,
which do not have to be shallow. Again, we partition $P$ into $\Theta(n/k)$ pairwise disjoint subsets, each having between $k$ and $2k$ points (here $1 < k
< n$ is an arbitrary parameter), enclosed by some simplex, where these simplices might have a nonempty intersection \footnote{A recent refined construction
by Chan[XX] constructs simplices with pairwise disjoint simplices.}. The partition, referred to as a \emph{simplicial partition}, has the property, that any
hyperplane $h$ crosses at most $O((n/k)^{1-1/d})$ simplices.

Let $P$, $n$ and $k$ be as above. Matou\v{s}ek \cite{mat92b} has also developed a variant of the partitioning scheme described above, yielding a partition
of $\Theta(n/k)$ subsets of size between $k$ and $2k$, such that any $k$-{\em shallow} hyperplane $h$ (that is, a hyperplane which contains at most $k$
points of $P$ on one side) crosses at most $O((n/k)^{1-1/\lfloor d/2 \rfloor})$ simplices of the partition.

(In contrast, the construction in \cite{hs11}, and the one presented in this paper, deal with sets whose points are shallow, but we seek a small crossing
number with respect to every hyperplane.)


\paragraph{Applications: related work.} Let us now review some applications based on the various kinds of partitions mentioned above.

\paragraph{Halfspace range searching.} Matou\v{s}ek \cite{mat92a} uses the standard partition recursively to construct an efficient {\em partition tree},
whose root stores the entire input set $P$ and a simplicial partition thereof. Each node $v$ of the tree stores a subset $P_{v}$ of $P$, and a simplicial
partition of $P_{v}$, and each subset of the partition corresponds to a distinct child of $v$. Overall, the resulting partition tree has linear size, can be
constructed in $O(n \log n)$ deterministic time, such that, given a query halfspace $\gamma$, one can count the number of points in $P \cap \gamma $ in
$O(n^{1-1/d} \log^{O(1)}n)$ time.

In a variant of this approach, Matou\v{s}ek \cite{mat92b} exploits the partition machinery with respect to shallow hyperplanes to efficiently solve the
halfspace range reporting (or emptiness) problem. More specifically, one can construct a data structure of size $O(n \log \log n)$, in $O(n \log n)$ time,
such that, given a query halfspace $\gamma$, one can report the points in $P \cap \gamma$ in $O(n^{1-1/\lfloor d/2 \rfloor}\log^{O(1)}n + k)$ time,
where $k = |P \cap \gamma|$.

Another useful application of partitions of sets of shallow points, as described above, is the construction of spanning trees with small {\em relative}
crossing number. We recall the standard result, due to Chazelle and Welzl \cite{cw89} (see also \cite{wel92}), on the existence of spanning trees with small
crossing number. That is, given a set $P$ of $n$ points in $\reals^{d}$, there exists a straight-edge spanning tree $\Tree$ on $P$, such any hyperplane
crosses at most $O(n^{1-1/d})$ edges of $\Tree$. Har-Peled and Sharir have refined this construction in the plane, to obtain a spanning tree $\Tree$ that
has the following property. Define the {\em weight} $w_{l} \leq n/2$ of a line $l$ to be the smaller of the two numbers of points of $P$ on each side of
$l$. Then $l$ crosses only $O(\sqrt{w_{l}}\log(n/w_{l}))$ edges of $\Tree$; see \cite{hs11} for more details. In this paper, we follow the same machinery to
extend the above construction to higher dimensions.

\paragraph{Relative $\peps$-approximations.} The existence of a spanning tree with small relative crossing number, for a point set $P$ in the plane, as just
reviewed, facilitates the construction of {\em relative $\peps$-approximations} for $P$, with respect to halfplane ranges, whose size is smaller than the
one guaranteed by the general theory of Li et al. \cite{lls01} (see \cite{hs11}). For given $0 < p,\eps < 1$, a relative $\peps$-approximation for $P$ is a
subset $A \subseteq P$ with the property that, for each halfspace $\gamma$,
$$
(1-\eps)\frac{|A \cap \gamma|}{|A|} \leq \frac{|P \cap \gamma|}{|P|} \leq
(1+\eps)\frac{|A \cap \gamma|}{|A|},
$$
provided that $|P \cap \gamma|/|P| \geq p$. Given such a spanning tree, we convert it into a spanning path with the same asymptotic crossing number, and
generate a matching out of this path by selecting every other edge along it, and apply a ``halving technique'' on the resulting matching, in which one point
of each pair of the matching is drawn independently at random, thus getting rid of half of the points. By the standard theory of discrepancy (see
\cite{cha01}), the small crossing number of the matching implies a correspondingly low discrepancy of the halving. We repeat this halving procedure until
the combined discrepancy first exceeds a given prescribed parameter, and then return the remaining points as the desired approximation. Har-Peled and Sharir
\cite{hs11} apply this construction to a planar point set $P$, and obtain a relative $\peps$-approximation for $P$, of size
$$
O\left(\frac{1}{\eps^{4/3}p}\log^{4/3}\frac{1}{\eps p}\right).
$$
For most values of $\eps$ and $p$, this is better than the general bound $O(\tfrac{1}{\eps^{2}p}\log\tfrac{1}{p})$ given in \cite{lls01}. See Section \ref{subsection_approximations} below for more details. In this paper, we extend this construction to higher dimensions, using the extension of spanning trees with small relative crossing number to higher dimensions, as mentioned above.


\paragraph{Our results.} We introduce a new variant of the partitioning machinery, based on the general approach that was introduced in \cite{mat92a} and
\cite{mat92b}, and obtain a partition of a set $P$ of $n$ $k$-shallow points in $\reals^{d}$ into $\Theta(n/k)$ subsets, each of size between $k$ and $2k$,
so that each subset is contained within a $d$-dimensional simplex and so that the crossing number of the partition (the maximum number of simplices crossed
by a hyperplane) is at most $O((n/k)^{1-1/(d-1)} \log^{2/(d-1)}(n/k))$ for $d \geq 3$ and $O(\alpha(n/k)\log^{2}(n/k))$ for $d = 2$, where $\alpha(\cdot)$
is the near-constant inverse Ackermann function. Note that the size of the sets in the partition is roughly the same as the shallowness parameter $k$, and
that the exponent in the bound on the crossing number is smaller than the one provided for the general setup in \cite{mat92a}.

We use this partition, as in \cite{hs11}, to construct a spanning tree with small relative, output-sensitive, crossing number.
Specifically, any hyperplane $h$ of weight $w_{h}$ (the number of input points on its ``lighter" side) crosses at most $O(n^{1-1/(d-1)}
w_{h}^{1/d(d-1)}\log^{2/(d-1)}(n/w_{h}))$ edges of the resulting spanning tree for $d \geq 3$, and $O(\sqrt{w_{h}} \alpha(n/w_{h}) \log^{2}(n/w_{h}) +
\alpha(n)\log^{4}n)$ edges for $d = 2$.

We then extend the planar construction of relative $\peps$-approximations for points and halfplanes introduced in \cite{hs11} to higher dimensions. We again
follow a similar machinery as the one in \cite{hs11}, based on the classical halving technique, to convert a spanning tree with small relative crossing
number to a relative $\peps$-approximation. The properties of the relative $\peps$-approximation and the exact result are detailed in Section
\ref{subsection_approximations} below.

Another application of the shallow-points partition is an exact output-sensitive range counting algorithm for point sets and halfspace ranges in
$\reals^{d}$. We combine the range reporting algorithm of Matou\v{s}ek \cite{mat92b} with our shallow-points partition and the standard simplex range
counting algorithm of Matou\v{s}ek \cite{mat92a}, to obtain an improved, output-sensitive range counting algorithm, whose running time is $O(n^{1-1/(d-1)}
k^{1/d(d-1)} (\log n)^{O(1)})$, for $d \geq 3$, where $k$ is the output size. This is an improvement over the previous bound $O(n^{1-1/d} (\log n)^{O(1)})$
of \cite{mat92a} when $k \ll n$ (although when $k$ is very small, the range reporting of \cite{mat92b} will be faster). See Section
\ref{subsection_range_countning} for a more precise statement of this result. (One weak aspect of our structure is that, for $d \geq 4$ we do not know how
to construct it in near-linear time. The current bound on the preprocessing cost is roughly $n^{c_{d}}$ for some exponent $c_{d}$ that depends on $d$ and
satisfies $1 < c_{d} < 3/2$. See Theorem \ref{thm_range_counting} for the precise statement.)

\subsection{Known notions of approximations} \label{subsection_approximations}
We recall the result of Li et al. \cite{lls01}, and two useful extensions of $\eps$-approximations and $\eps$-nets derived from it, which we will exploit later in this paper.

Let $(X,\R)$ be a \emph{range space}, where $X$ is a set of $n$ objects and $\R$ is a collection of subsets of $X$, called \emph{ranges}. The \emph{measure} of a range $\r \in \R$ in $X$ is the quantity
$$
\meas{X}{\r} = \frac{|X \cap \r|}{|X|}.
$$
Assume that $(X,\R)$ has {\em finite VC-dimension} $\delta$, which is a constant independent of $n$; see \cite{mat02} for more details.

Let $0< \alpha,\nu <1$ be two given parameters. Consider the distance function
$$
d_\nu(r,s) = \frac{|r-s|}{r+s+\nu}, \quad\quad \mbox{for $r,s \geq 0$} .
$$
A subset $N \subseteq X$ is called a {\em $(\nu, \alpha)$-sample} for $(X,\R)$, if for each $\r \in \R$ we have
$$
d_\nu\left(\meas{X}{\r}, \meas{N}{\r} \right) < \alpha .
$$

\begin{theorem} \label{thm_nu_alpha_sample} $(${\emph{Li et al.} \bf \cite{lls01}}$)$
A random sample of $X$ of size
$$
O\left(\frac{1}{\alpha^2\nu} \left(\delta \log\frac{1}{\nu} + \log\frac{1}{q} \right) \right)
$$
is a $(\nu,\alpha)$-sample for $(X,\R)$ with probability at least $1-q$, with an appropriate choice of an absolute constant of proportionality.
\end{theorem}

Har-Peled and Sharir \cite{hs11} show that, by appropriately choosing $\alpha$ and $\nu$, various standard constructs, such as $\eps$-nets and $\eps$-approximations, are special cases of $(\nu,\alpha)$-samples; thus bounds on the sample sizes that guarantee that the sample will be one of these constructs with high probability are immediately obtained from Theorem \ref{thm_nu_alpha_sample}. Two other, less standard special cases of $(\nu,\alpha)$-samples are relative $\peps$-approximations \cite{hs11} and {\em shallow $\eps$-nets} \cite{shsh11}; we review them next.

\paragraph{Relative $\peps$-approximations.}

\begin{defn} \label{definition_peps}
$(${\em Relative $\peps$-Approximation}$)$. Let $(X,\R)$ be a range space of finite VC-dimension $\delta$. For given parameters $0 < p,\eps < 1$, a subset $N \subseteq X$ is a \emph{relative $\peps$-approximation} for $(X,\R)$ if, for each $\r \in \R$, we have
\begin{center}
\begin{tabular}{lp{1\linewidth}}
(i) $(1-\eps) \meas{X}{\r} \leq \meas{N}{\r} \leq (1+\eps)\meas{X}{\r}$, \quad & if  $\meas{X}{\r} \geq p.$ \\[5pt]
(ii) $\meas{X}{\r} - \eps p \leq \meas{N}{\r} \leq \meas{X}{\r} + \eps p$,  \quad & if $\meas{X}{\r} \leq p.$
\end{tabular}
\end{center}
\end{defn}

The following result shows that a relative $\peps$-approximation is a special case of a $(\nu,\alpha)$-sample.

\begin{theorem} \label{thm_relative_peps} $(${\emph{Har-Peled and Sharir} \bf \cite{hs11}}$)$ A random sample $N$ of
$$
\frac{c_{1}}{\eps^{2}p}\left(\delta \log \frac{1}{p} + \log \frac{1}{q} \right)
$$
elements of $X$, for an appropriate absolute constant $c_{1}$, is a relative $\peps$-approximation for $(X,\R)$ with probability at least $1-q$.
\end{theorem}

\paragraph{Shallow $\eps$-nets.}

Let $(X,\R)$ be a range space of finite VC-dimension $\delta$, and let $0 < \eps < 1$ be a given parameter. A subset $N \subseteq X$ is a \emph{shallow $\eps$-net} if it satisfies the following two properties, for some constant $c$ that depends on $\delta$.

\begin{list}{(\roman{itemcounter})}{\usecounter{itemcounter} \leftmargin=2em}
\item For each $\r \in \R$ and for any parameter $t \geq 0$, if $|N \cap \r| \leq t\log\frac{1}{\eps}$ then $|X \cap \r| \leq c(t+1)\eps|X|$.
\item For each $\r \in \R$ and for any parameter $t \geq 0$, if $|X \cap \r| \leq t\eps|X|$ then
    $|N \cap \r| \leq c(t+1) \log \frac{1}{\eps}$.
\end{list}

Note the difference between shallow and standard $\eps$-nets: Property (i) (with $t=0$) implies that a shallow $\eps$-net is also a standard $\eps$-net (possibly with a re-calibration of
$\eps$). Property (ii) has no parallel in the case of standard $\eps$-nets --- there is no guarantee how a standard $\eps$-net interacts with small ranges.

\begin{theorem} \label{epsshallow} $(${\emph{Sharir and Shaul} \bf \cite{shsh11}}$)$
A random sample $N$ of
$$
\frac{c'_{1}}{\eps} \left( \delta \log\frac{1}{\eps} +
\log\frac{1}{q} \right)
$$
elements of $X$, for an appropriate absolute constant $c'_{1}$, is a shallow $\eps$-net with probability at least
$1-q$.
\end{theorem}

\paragraph{A brief overview of the analysis.}

The paper is structured as follows. In Section \ref{section_sampling} we derive a technical result that shows that, informally, a random sample $Z$ of $\Theta(n/k)$ points from a $k$-shallow $n$-point set in $\reals^{d}$ has, with some constant positive probability, the property that at least some fixed fraction of its points are vertices of its convex hull. (Curiously, this fraction cannot exceed $1/e \approx 0.368$ in the worst case.) We plug this result into the analysis in Section \ref{section_cutting_lemma}, which presents a construction of a $(1/r)$-cutting of the zone of the boundary of a convex set in an arrangement of hyperplanes in $\reals^{d}$. This construction is an easy adaption of standard constructions of cuttings, but seems not to have been presented so far in the literature. Armed with this construct, we present in Section \ref{section_partition} the main technical contribution of the paper, which is a construction of a simplicial partition of a set of shallow points in $\reals^{d}$ that has a small crossing number (with respect to any hyperplane). The construction is not general, in the sense that the size of the subsets in the partition has to be equal to (or proportional to) the shallowness parameter $k$. Still, this provides the key tool for the applications in Section \ref{section_applications}, which, as mentioned above, are $(i)$ a construction of a spanning tree with small relative crossing number, $(ii)$ a construction of relative $\peps$-approximations for halfspace ranges in $\reals^{d}$, whose size (in many cases) is smaller than the halfspace general bound in \cite{lls01,hs11}, and $(iii)$ an output-sensitive halfspace range searching data structure.


\section{Sampling from a $k$-shallow set} \label{section_sampling}

\paragraph{}
Let $P$ be a set of $n$ points in $\reals^{d}$, for $d \geq 2$, so that all its points are $k$\emph{-shallow}, for some fixed parameter $1 \leq k \leq n$.

Let $Z$ be a random sample of $P$ where each point of $P$ is chosen in $Z$ independently with probability $p = 1/k$. Let $Y\subseteq Z$ be the set of $1$-shallow points of $Z$ (with respect to $Z$), i.e., $Y$ is the set of vertices of the convex hull $Conv(Z)$. Let us assume from now on that $k \geq 2$, unless explicitly noted otherwise.

In what follows, we assume that $P$ is in general position. In particular, no $d+1$ points of $P$ lie in a common hyperplane, and no $d$ lie in a common vertical hyperplane (i.e., parallel to the $x_{d}$-axis). Consequently, considering the dual set of hyperplanes $H = \mathcal{D}(P)$ (using the duality that preserves the above/below relationship; see, e.g., \cite{ed87}), we get that $H$ is in general position too, which means that every $d$ hyperplanes of $H$ intersect in a single point, and no $d+1$ have a common point.


The following proposition, although its proof is technical and somewhat involved, asserts a fairly intuitive property of $k$-shallow sets. That is, if we take a random sample of about $n/k$ points of such a set, at least some fixed fraction of the points of the sample are vertices of its convex hull. As a partial explanation of why the proof is complicated, we observe, after the proof, that the fraction of convex hull vertices of the sample cannot in general be made arbitrarily close to $1$.

\begin{proposition} \label{lemma_prob} Let $P$, $Z$, and $Y$ be as above, and assume that $k \leq \frac{n}{256}$. Then, with probability $> 0.22$,
$$
|Y| \geq \frac{n}{8k}, \quad \text{and} \quad \frac{n}{2k} \leq |Z| \leq \frac{2n}{k}.
$$
\end{proposition}

\paragraph{}
Before getting into the proof, let us recall Chernoff's bound (see, e.g., \cite{as92}, Appendix A, Theorems A.12 and A.13) for a binomial random variable $X$ with parameters $n$, $p$, which we denote by $X \sim B(n,p)$. That is, $X = \sum_{i=1}^{n} X_{i}$, where $X_{1}, X_{2}, \ldots, X_{n}$ are independent random indicator variables, and ${\bf Pr}\left\{ X_{i} = 1 \right\} = p$ for each $i$. Put ${\bf E}[X] = \mu = np$. Then, for any $\delta > 0$, we have
\begin{equation} \label{eq_chernoff1}
{\bf Pr}\left\{X > (1+\delta)\mu \right\} < \left(\frac{e^{\delta}}{(1+\delta)^{(1+\delta)}}\right)^{\mu} \equiv \eps(\mu,1+\delta).
\end{equation}
Alternatively, we can write it as
\begin{equation} \label{eq_chernoff2}
{\bf Pr}\left\{X > t \right\} < \left(\frac{e^{1-\frac{\mu}{t}}\mu}{t} \right)^{t} = \eps(\mu, t/\mu),
\end{equation}
for any $t > \mu$. Another variant of Chernoff's bound, for large deviations below the mean, is
\begin{equation} \label{eq_chernoff3}
{\bf Pr}\left\{X < (1 - \delta)\mu \right\} < e^{-\frac{\mu\delta^{2}}{2}},
\end{equation}
for any $\delta > 0$. Using $(\ref{eq_chernoff1})$ and $(\ref{eq_chernoff2})$, we get the following proposition:
\begin{proposition} \label{prop_sum}
Let $X \sim B(n,\frac{1}{k})$, and let $1 < \beta \leq k$ be a parameter. For $y \in \reals$, define
$$
S_{> y} = \sum_{t > y \ \mid \ t \in \mathbb N} t \cdot {\bf Pr}\left\{X = t\right\}, \quad\quad
S_{\geq y} = \sum_{t \geq y \ \mid \ t \in \mathbb N} t \cdot {\bf Pr}\left\{X = t\right\}.
$$
Then
$$
S_{> \frac{\beta n}{k}} \leq \eps(n/k, \beta) \cdot \left(\frac{\beta n}{k} + 1 + \frac{\beta}{\beta - e^{1-1/\beta}} \right).
$$
\end{proposition}

\noindent{\bf Proof.} We have, for $t_{0} \in \mathbb N$,
$$
S_{\geq t_{0}} = \sum_{t \geq t_{0}} t \cdot {\bf Pr}\left\{X = t \right\} =
\sum_{t \geq t_{0}} t \left({\bf Pr} \left\{X \geq t \right\} - {\bf Pr}\left\{X \geq t + 1 \right\} \right) =
$$
$$
t_{0} \cdot {\bf Pr}\left\{X \geq t_{0}\right\} + \sum_{t = t_{0}+1}^{n} {\bf Pr}\left\{X \geq t\right\} = t_{0} \cdot {\bf Pr}\left\{X \geq t_{0} \right\} + \sum_{t = t_{0}}^{n-1} {\bf Pr}\left\{X > t\right\}.
$$
We note, that for any $y \in \reals$, $S_{> y} = S_{\geq \lfloor y+1 \rfloor}$. Hence, substituting $t_{0} = \lfloor \frac{\beta n}{k} + 1 \rfloor$ into the equality above, and  using the bound $(\ref{eq_chernoff1})$, with $\mu = \frac{n}{k}$ and $1 + \delta = \beta$, we get
$$
S_{> \frac{\beta n}{k}} = S_{\geq \lfloor \frac{\beta n}{k}+1 \rfloor} =
\left\lfloor \frac{\beta n}{k}+1 \right\rfloor \cdot {\bf Pr}\left\{X \geq \left\lfloor \frac{\beta n}{k}+1 \right\rfloor \right\} + \sum_{t = \lfloor \frac{\beta n}{k}+1 \rfloor}^{n-1} {\bf Pr}\left\{X > t\right\} \leq
$$
$$
\left(\frac{\beta n}{k} + 1\right) {\bf Pr}\left\{X > \frac{\beta n}{k}\right\} + \sum_{t = \lfloor \frac{\beta n}{k} + 1 \rfloor}^{n-1}{\bf Pr}\left\{X > t \right\} \leq
$$
$$
\left(\frac{\beta n}{k} + 1\right)\cdot\eps(n/k, \beta) + \sum_{t = \lfloor \frac{\beta n}{k} + 1 \rfloor}^{n-1}{\bf Pr}\left\{X > t \right\}.
$$
It remains to estimate the second expression. We note that $(e^{1-a/x})/x$ is monotonically decreasing in $x$ for $0 < a \leq x$. Hence, using $(\ref{eq_chernoff2})$, we obtain, for any $y > \mu$,
$$
\sum_{t \geq y}{\bf Pr}\left\{X > t\right\} \leq \sum_{t \geq y}\left(\frac{e^{1-\frac{\mu}{t}} \mu}{t}\right)^{t} \leq \sum_{t \geq y}\left(\frac{e^{1-\frac{\mu}{y}}\mu}{y}\right)^{t} <
$$
$$
\left(\frac{e^{1-\frac{\mu}{y}}\mu}{y}\right)^{y}\left(\frac{y}{y - e^{1-\frac{\mu}{y}}\mu}\right) = \eps(\mu, y/\mu) \cdot \left(\frac{y}{y - e^{1-\frac{\mu}{y}}\mu}\right).
$$
Putting $y = \frac{\beta n}{k}$, and noting that $y > \mu$ (since $\beta > 1$ and $\mu = \frac{n}{k}$), we obtain
$$
\sum_{t = \lfloor \frac{\beta n}{k} + 1 \rfloor}^{n-1}{\bf Pr}\left\{X > t \right\} \leq \sum_{t \geq \frac{\beta n}{k}}{\bf Pr}\left\{X > t \right\} \leq \eps(n/k,\beta) \cdot \left(\frac{\beta}{\beta - e^{1-1/\beta}}\right),
$$
which completes the proof. $\Box$

\noindent{\bf Proof of Proposition $\ref{lemma_prob}$.}
We separately bound from below the probabilities that (i) $|Y| \geq \frac{\alpha n}{k}$, and (ii) $\frac{n}{\nu k} \leq |Z| \leq \frac{\gamma n}{k}$, in terms of $\alpha$, $\nu$, $\gamma$, $n$ and $k$; the concrete values of $\alpha$, $\nu$ and $\gamma$, as given in the proposition, will be substituted later. The probability for both conditions to hold is then estimated using the union bound (on the complementary events).

\paragraph{(i) Bounding from below the probability that $|Y| \geq \frac{\alpha n}{k}$.}
Put $q = {\bf Pr}\left\{|Y| < \frac{\alpha n}{k} \right\}$, and let $1 < \beta \leq k$ be a parameter. We need to bound from below the value of $1-q$.

Since $Y \subseteq Z$ and $|Z| \sim B(n,\frac{1}{k})$, $(\ref{eq_chernoff1})$ implies that
\begin{equation} \label{eq1}
{\bf Pr}\left\{|Y| > \frac{\beta n}{k}\right\} \leq  {\bf Pr}\left\{|Z| > \frac{\beta n}{k}\right\} < \eps(n/k, \beta).
\end{equation}
Recalling that $Y$ is the subset of $1$-shallow points within $Z$, and that all the points in the ground set $P$ are $k$-shallow, we have, for each point $a\in P$,
$$ {\bf Pr}\left\{a \in Y \right\} \geq \frac{1}{k}\left(1-\frac{1}{k}\right)^{k-1} \geq \frac{1}{ke}. $$
Indeed, as $a$ is a $k$-shallow point, there exists a hyperplane through $a$ which bounds a halfspace $h$ containing at most $k-1$ points of $P$, excluding $a$. The lower bound above is the probability that $a$ is chosen in $Z$ and none of the other points in $h$ is chosen. Using linearity of expectation, we thus have

\begin{equation}\label{eq2}
{\bf E}\{|Y|\} \geq \frac{|P|}{ke} = \frac{n}{ke}.
\end{equation}

We now proceed to estimate $q$. Note that we cannot apply Chernoff's bound directly, because the events $a \in Y$, for $a \in P$, are not independent. Instead we proceed as follows. Put $\eps = \eps(n/k,\beta)$, and  $s = S_{> \frac{\beta n}{k}}$. We then have (assuming that $\beta > \alpha$, which will hold for specific values that we will later use)
\begin{align*}
\frac{n}{ke} \leq {\bf E}\{|Y|\} =& \sum_{t=0}^n t\cdot {\bf Pr}\left\{|Y|=t\right\} = \\
&\sum_{t < \frac{\alpha n}{k}} t\cdot {\bf Pr}\left\{|Y|=t \right\} +
\sum_{\frac{\alpha n}{k} \leq t \leq \frac{\beta n}{k}} t\cdot {\bf Pr}\left\{|Y|=t\right\} + \\
&\sum_{t > \frac{\beta n}{k}} t\cdot {\bf Pr}\left\{|Y|=t \right\} \leq
q\frac{\alpha n}{k} + (1-q-\eps)\frac{\beta n}{k} + s.
\end{align*}
Hence, by applying Proposition $\ref{prop_sum}$, we have
$$
q \leq \frac{\beta(1-\eps) - e^{-1} + s\frac{k}{n}}{\beta-\alpha} \leq
\frac{\beta - e^{-1} + \eps\frac{k}{n}\left(1 + \frac{\beta}{\beta - e^{1-1/\beta}}\right)}{\beta-\alpha},
$$
or
\begin{equation} \label{eq_sample_k}
1-q \geq 1 - \frac{\beta - e^{-1} + \eps\frac{k}{n}\left(1 + \frac{\beta}{\beta - e^{1-1/\beta}}\right)}{\beta-\alpha}.
\end{equation}
By choosing (somewhat arbitrarily) $\alpha = \frac{1}{8}$, $\beta = 1.2$, recalling that we have required that $k \leq \frac{n}{256}$, and using Chernoff's bound (\ref{eq_chernoff1}) to bound $\eps = \eps(n/k,\beta)$, we get that ${\bf Pr}\left\{|Y| \geq \frac{n}{8k}\right\} > 0.221$.

\paragraph{(ii) Bounding from below the probability that $\frac{n}{\nu k} \leq |Z| \leq \frac{\gamma n}{k}$.} Since $|Z| \sim B(n,\frac{1}{k})$, estimating this probability can be done by a direct application of Chernoff's bounds. As in (\ref{eq1}),
$$
{\bf Pr}\left\{|Z| > \frac{\gamma n}{k} \right\} < \eps(n/k,\gamma),
$$
and by applying (\ref{eq_chernoff3}), we have
$$
{\bf Pr}\left\{|Z| < \frac{n}{\nu k} \right\} < e^{-\frac{n}{k}\frac{(1-1/\nu)^{2}}{2}}.
$$
Choosing $\gamma = 2$, $\nu = 2$, and using the assumption that $k \leq \frac{n}{256}$, we get that $$
{\bf Pr}\left\{\frac{n}{2k} \leq |Z| \leq \frac{2n}{k} \right\} > 0.999.
$$
From (i) and (ii) we obtain that
$$
{\bf Pr}\left\{\left(|Y| \geq \frac{n}{8k}\right) \wedge \left(\frac{n}{2k} \leq |Z| \leq \frac{2n}{k} \right) \right\} > 0.22,
$$
as asserted. $\Box$


\paragraph{Upper bounds on the probability of a sample with many hull vertices.}
We can increase the lower bound given in (\ref{eq_sample_k}), on the probability $1-q$ of getting a sample with at least $\frac{\alpha n}{k}$ hull vertices, by choosing $\beta$ close enough to $1$ and $\alpha$ close to $0$. We then have
$$
1-q \geq 1 - \frac{\beta - e^{-1} + \eps\frac{k}{n}\left(1 + \frac{\beta}{\beta - e^{1-1/\beta}}\right)}{\beta-\alpha}
\approx 1 - \left(1 - e^{-1} + \eps\frac{k}{n}\left(1 + \frac{\beta}{\beta - e^{1-1/\beta}}\right)\right).
$$
By requiring that $n/k$ is at least some sufficiently large constant (which depends on the choice of $\beta$; we need it to "neutralize" the denominator $\beta - e^{1-1/\beta}$, which tends to $0$ as $\beta \rightarrow 1$), we can make this probability arbitrarily close to
$$
1 - (1 - e^{-1}) = e^{-1} \approx 0.368.
$$

The following construction shows that the bound $1/e$ is fairly close to being worst-case tight. This construction was provided by Adam Sheffer, and we are grateful to him for this observation.

\begin{figure}[htb]
    \begin{center}
        \scalebox{0.65}{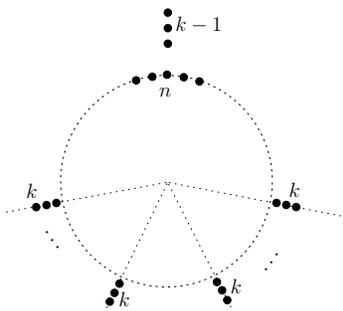}
        \caption{Construction of a $k$-shallow point set with large probability for a sample with a constant number of hull vertices.}
        \label{fig_sample}
    \end{center}
\end{figure}

The construction is depicted in Figure \ref{fig_sample}. It consists of a set $A$ of $n$ points on the unit circle $C$, sufficiently closely clustered together near the top point of $C$, and of $2c+1$ sets $T$, $L_1,\ldots,L_c$, and $R_1,\ldots,R_c$, so that (i) each of these sets consists of points lying on a ray emanating from the center of $C$, outside $C$ and sufficiently close to it, (ii) each of these sets consists of $k$ points, except for $T$ which consists of $k-1$ points, and (iii) the ray containing $T$ points upwards, the rays containing
$L_1,\ldots,L_c$ point into the third quadrant, and the rays containing $R_1,\ldots,R_c$ point into the fourth quadrant. Here $c$ is an additional parameter that we will fix later. (The points are not in convex position but a sufficiently small perturbation of them will place them in general position without affecting the analysis.)

Altogether, the resulting set $P$ has $n'=n+(2c+1)k-1$ points. It is easily verified that the points can be arranged in such a way that each of them is $k$-shallow. Moreover, with an appropriate layout, the convex hull of a sample $Z$ will have at most $2c+1$ vertices if we choose at least one point of $T$, at least one point of $L_1\cup\cdots\cup L_c$, and at least one point of $R_1\cup\cdots\cup R_c$. The probability of this latter event is
$$
q' = \left(1 - \left(1-\frac{1}{k}\right)^{k-1}\right)
\left(1 - \left(1-\frac{1}{k}\right)^{ck}\right)^2 \approx
\left(1-e^{-1}\right) \left(1-e^{-c}\right)^2 .
$$
Assuming that $n \gg (2c+1)k$, so that $2c+1 \ll n'/k$, the probability of getting a sample with $\Theta(n'/k)$ hull vertices, with an appropriate constant of proportionality ensuring that this bound is larger than $2c+1$, is
$$
1 - q' = 1 - \left(1-e^{-1}\right) \left(1-e^{-c}\right)^2 ,
$$
which we can make arbitrarily close to
$$
1 - \left(1-e^{-1}\right) = e^{-1},
$$
by choosing $c$ sufficiently large. This shows that $e^{-1}$ is an upper bound on the probability that we can guarantee for getting a sample with $\Theta(n/k)$ hull vertices from a set of $n$ $k$-shallow points (when $n$ is sufficiently large). As noted above, we can get arbitrarily close to this bound by requiring $k/n$ to be sufficiently small.

\paragraph{$Z$ is also a relative approximation set.}

Consider next the range space $(P,\R)$, where $\R$ is the collection of all simplices in $\reals^{d}$. As is well known,  $(P,\R)$ has finite VC-dimension, which we denote by $\delta$. As a matter of fact, we have $\delta = O(d^{2}\log d)$; see, e.g., [SA95??].

Proposition \ref{lemma_prob} establishes a property of a random sample of size roughly $n/k$ from a set of $n$ $k$-shallow points in $\reals^{d}$. Of course, such a sample is also a $(\nu,\alpha)$-sample with high probability, for appropriate choices of $\nu$ and $\alpha$. The following proposition combines both properties.

\noindent{\bf Remark:} The standard sampling method, under which Theorem \ref{thm_relative_peps} was originally proved, is to choose each subset of the prescribed size $s$ with equal probability. However, the theorem also holds if the subset is obtained by choosing each point $x \in X$ independently with probability $s/|X|$, similar to the way $Z$ was chosen. In this case $s$ is only the expected size of the sample.

\begin{proposition} \label{prop_peps}
Let $P$, $Z$, $n$, $k$ be as above, and let $c_{1}$ be the constant in Theorem \ref{thm_relative_peps}. Then, with probability at least $q = 0.999$, $Z$ is a relative $(\frac{c_{2}k}{n}\log\frac{n}{c_{2}k},1/2)$-approximation for $(P,\R)$, for $c_{2} \geq 8 c_{1} (\delta + \log (1/q))$, and for $n \geq 2c_{2}k$.
\end{proposition}

\noindent{\bf Proof.} As already noted, the sample $Z$ satisfies $|Z| \geq \frac{n}{2k}$, it suffices to show that
$$
\frac{c_{1}}{\eps^{2}p}\left(\delta \log \frac{1}{p} + \log \frac{1}{q} \right)  \leq \frac{n}{2k},
$$
for the specific values $p = \frac{c_{2}k}{n}\log\frac{n}{c_{2}k}$, $\eps = 1/2$, $q = 0.999$, and $c_{2}$ as set above. Indeed, we have
$$
\frac{c_{1}}{\eps^{2}p}\left(\delta \log \frac{1}{p} + \log \frac{1}{q} \right) \leq
\frac{4 c_{1} \delta}{c_{2}} \left(1 - \frac{\log \log \frac{n}{c_{2}k}}{\log \frac{n}{c_{2} k}}\right) \cdot \frac{n}{k} + \frac{4 c_{1}}{c_{2}} \cdot \frac{\log \frac{1}{q}}{\log \frac{n}{c_{2} k}} \cdot \frac{n}{k} \leq
$$
$$
\frac{4 c_{1}(\delta + \log \frac{1}{q})}{c_{2}} \cdot \frac{n}{k} \leq \frac{n}{2k},
$$
as asserted, provided that $n \geq 2 c_{2} k$ (which implies that $\log \frac{n}{c_{2} k} \geq 1$). $\Box$


\section{Cutting lemma for the zone of a convex set} \label{section_cutting_lemma}

\paragraph{}
Let $H$ be a collection of $n$ hyperplanes in $\reals^d$. We denote the arrangement of $H$ by $\mathcal{A}(H)$. For a simplex $\Delta$, let $H_{\Delta}$ denote the collection of hyperplanes of $H$ that {\em cross} $\Delta$ (i.e., intersect its interior). A $(1/r)$\emph{-cutting} for $\mathcal{A}(H)$ is a collection $\Xi$ of (possibly unbounded) closed $d$-dimensional simplices with pairwise disjoint interiors, which cover $\reals^{d}$, such that $|H_{\Delta}| \leq n/r$, for every $\Delta \in \Xi$. The \emph{size} of a $(1/r)$-cutting is the number of its simplices. The so-called Cutting Lemma (see, e.g., \cite{mat02}) asserts that for every $H$ and $r \leq n$ there exists a $(1/r)$-cutting of size $O(r^{d})$ for $H$ (which is asymptotically the best possible size).

Here we will need a modified version of the Cutting Lemma, where the cutting is not required to cover the entire $\reals^{d}$, but only the zone in $\mathcal{A}(H)$ of the boundary $\partial C$ of some fixed convex region $C$, where the zone of $\partial C$ is the collection of all cells of $\mathcal{A}(H)$ whose interiors are crossed by $\partial C$. The simplices of our cutting might also contain points outside this zone, but they are required to cover all the cells of the zone. (In a sense, this can be regarded as a variant of the shallow cutting lemma of Matou\v{s}ek \cite{mat92b}.)

\begin{theorem} \label{thm_cutting}
\emph{({\bf Cutting Lemma for the zone of a convex set})}. Let $H$ be a collection of $n$ hyperplanes in $\reals^d$, let $r \leq n$ be a parameter, and let $\sigma$ be the boundary of a convex set in $\reals^d$. Then there exists a $(1/r)$-cutting $\Xi$ for the zone of $\sigma$ in the arrangement $\mathcal{A}(H)$, consisting of $O(r^{d-1}\log r)$ simplices, for $d \geq 3$, and of $O(r \alpha(r))$ simplices, for $d = 2$.
\end{theorem}

We will need a variant of this result for weighted collections of hyperplanes, where such a collection is a pair $(H,w)$, where $H$ is a collection of hyperplanes, and $w:H \rightarrow \reals^{+}$ is a nonnegative weight function on $H$. For a subset $X \subseteq H$, we write $w(X)$ for $\sum_{h \in X}w(h)$. The notions introduced for unweighted collections of hyperplanes can usually be generalized for weighted collections in an obvious way. In particular, given the boundary $\sigma$ of a convex set in $\reals^{d}$, we say that $\Xi$ is a $(1/r)$-cutting for the zone of $\sigma$ in $\mathcal{A}(H,w)$, if the simplices of $\Xi$ cover all points in the zone, and for every simplex $\Delta$ of $\Xi$, the collection $H_{\Delta}$ has total weight $w(H_{\Delta}) \leq w(H)/r$. A simple reduction from cuttings for weighted collections of hyperplanes to unweighted cuttings is discussed in \cite{mat91a}, and can be applied here for a $(1/r)$-cutting for the zone of $\sigma$ in $\mathcal{A}(H)$ without any change. Hence, Theorem \ref{thm_cutting} implies the following.

\begin{corollary} \label{corollary_weight_cutting}
Let $(H,w)$ be a weighted collection of $n$ hyperplanes in $\reals^{d}$, let $r \leq n$ be a parameter, and let $\sigma$ be the boundary of a convex set. There exists a $(1/r)$-cutting $\Xi$ for the zone of $\sigma$ in $\mathcal{A}(H,w)$ of size $O(r^{d-1}\log r)$, for $d \geq 3$, and $O(r \alpha(r))$, for $d = 2$.
\end{corollary}

Before proving Theorem \ref{thm_cutting}, we need the following technical lemma.
\begin{lemma} \label{lemma_func_expectation}
Let $X \sim B(n,p)$ and put $\mu = {\bf E}\{X\} = np$. Then
\begin{list} {\emph{(\roman{itemcounter})}}{\usecounter{itemcounter} \leftmargin=1em}
\item ${\bf E}\{X^{k}\log(X+1)\} = O(\mu^{k}\log \mu)$, for any $k \geq 1$.
\item ${\bf E}\{X \alpha(X+1)\} = O(\mu \alpha(\mu))$.
\end{list}
\end{lemma}
\noindent{\bf Proof.} This follows from the fact that binomial variables are concentrated near their means, and we omit the routine and technical details. $\Box$

\paragraph{}
The proof of Theorem \ref{thm_cutting} uses a triangulation of the cells of an arrangement of hyperplanes, called \emph{canonical triangulation} (or \emph{bottom-vertex triangulation}). The definition and some properties of this triangulation can be found in \cite{mat02}. For a subcollection $L \subseteq H$ of hyperplanes, let $CT(L)$ denote the set of simplices in the canonical triangulation of $\mathcal{A}(L)$. Let $\sigma$ be the boundary of a convex set in $\reals^{d}$. We denote by $CT_{\sigma}(L) \subseteq CT(L)$ the minimal set of simplices of $CT(L)$ which cover the zone of $\sigma$ in $\mathcal{A}(L)$. We recall the following results about the canonical triangulation:

\begin{lemma} \label{lemma_triangulation} Let $H$ be a collection of $n$ hyperplanes in $\reals^{d}$.
\begin{list} {\emph{(\roman{itemcounter})}}{\usecounter{itemcounter} \leftmargin=1em}
\item \emph{\cite{cf90}} For every simplex $\Delta \in CT(H)$, there exists a unique inclusion-minimal collection $S_{\Delta} \subseteq H$, such that $\Delta \in CT(S_{\Delta})$. This collection $S_{\Delta}$, called the \emph{defining set} of $\Delta$, consists of at most $D = d(d+3)/2$ hyperplanes.
\item \emph{\cite{cf90}} If $L \subseteq H$ and $\Delta \in CT(L)$ then $\Delta \in CT(H)$ if and only if its interior is intersected by no hyperplane of $H$.
\item \emph{\cite{mat02}} Assuming $H$ is in general position, each cell in $\mathcal{A}(H)$ is a simple polytope, in which the number of simplices in its canonical triangulation is at most proportional to the number of its vertices (with the constant of proportionality depending on $d$).
\end{list}
\end{lemma}

\paragraph{}

Let $H$, $n$, $r$ and $\sigma$ be as in Theorem \ref{thm_cutting}. For a simplex $\Delta$, define the \emph{excess} of $\Delta$ to be max$\{1, |H_{\Delta}|\frac{r}{n}\}$. Let $R$ be a random sample of hyperplanes of $H$, where each hyperplane is drawn independently with probability $p = r/n$. Let $n_{\sigma}(p,t)$ denote the expected number of simplices with excess at least $t$ in $CT_{\sigma}(R)$.

\begin{lemma}\label{npt_lemma}
\emph{{\bf(Exponential Decay Lemma)}}. For $t > 1$ and $r \leq n/2$,
$$n_{\sigma}(p, t) = O(2^{-t}n_{\sigma}(p/t, 1)),$$
\end{lemma}
where $n_{\sigma}(p/t, 1)$ is the expected size of $CT_{\sigma}(R')$, for another random sample $R'$, in which each hyperplane is chosen independently with probability $p/t$.

We omit the proof of the lemma, which is essentially the same as in \cite{mat92a} and \cite{mat92b}, and it can be viewed as a special instance of the general setup considered in \cite{ams98}.

\paragraph{}
As a final step before the proof of Theorem $\ref{thm_cutting}$, we recall the bound established in \cite{aps93} on the complexity of the zone of the boundary of a convex set in a hyperplane arrangement.

\begin{theorem}\label{thm_extended_zone}
\emph{{\bf(Extended Zone Theorem)}}. The complexity of the zone of the boundary of an arbitrary convex set in an arrangement of $n$ hyperplanes in $\reals^{d}$ is $O(n^{d-1}\log n)$ for $d \geq 3$, where the constant of proportionality depends on $d$, and $O(n\alpha(n))$ for $d = 2$.
\end{theorem}

\noindent{\bf Proof of Theorem \ref{thm_cutting}.} First, we may assume that $r \leq n/2$, since otherwise, the bottom-vertex triangulation $CT_{\sigma}(H)$ can serve as the desired cutting. By Lemma \ref{lemma_triangulation} (iii),  the number of simplices in $CT_{\sigma}(R)$ is at most proportional to the number of vertices in the zone of $\sigma$ in $\mathcal{A}(R)$. Hence, by Theorem \ref{thm_extended_zone}, we have


$$
|CT_{\sigma}(R)| =
\begin{cases}
    O(|R|^{d-1}\log(|R|+1)), & \mbox{for $d \geq 3$} \\
    O(|R|\alpha(|R|+1)), & \mbox{for $d=2$}.
\end{cases}
$$
Since $|R|\sim B(n,p)$, Lemma \ref{lemma_func_expectation} (i) implies that the expected number $n_{\sigma}(p,1)$ of simplices in $CT_{\sigma}(R)$ is, for $d \geq 3$,
$$
n_{\sigma}(p,1) = {\bf E}\{|CT_{\sigma}(R)|\} = {\bf E} \left\{O\left(|R|^{d-1}\log(|R|+1)\right)\right\} =
$$
\begin{equation}\label{eq_CT_R}
O(({\bf E}\{|R|\})^{d-1}\log({\bf E}\{|R|\})) = O((np)^{d-1}\log (np)).
\end{equation}
For $d = 2$, a similar argument shows, by Lemma \ref{lemma_func_expectation} (ii), that $n_{\sigma}(p,1) = O(np \alpha(np))$. Note that these bounds hold for any $0 < p \leq 1$.

To obtain the desired cutting, we start with $\Xi_{0} = CT_{\sigma}(R)$. For each simplex $\Delta \in \Xi_{0}$, let $t = t(\Delta)$ be the excess of $\Delta$. Each simplex $\Delta$ with $t(\Delta) = 1$ is left as is. For those simplices $\Delta$ with $t(\Delta) > 1$, let $\Xi_{\Delta}$ be a $(1/t)$-cutting for $\mathcal{A}(H_{\Delta})$ of size $O(t^{d})$. Such a cutting exists by the Cutting Lemma \cite{mat02}. We take the intersection of every simplex $\Delta ' \in \Xi_{\Delta}$ with $\Delta$, and triangulate it if necessary into $O(1)$ sub-simplices. The collection of simplices appearing in these triangulations, over all simplices $\Delta \in \Xi_{0}$, with $t(\Delta) > 1$, plus the simplices of $\Xi_{0}$ which were not triangulated further, form our cutting $\Xi$, which, as is easily verified, is indeed a $(1/r)$-cutting for the zone of $\sigma$ in $\mathcal{A}(H)$. We bound the size of $\Xi$ by bounding the expected value $S$ of the sum $\sum_{\Delta \in \Xi_{0}}t(\Delta)^{d}$. We have
$$
S \leq \sum_{t=1}^{\infty}t^{d}n_{\sigma}(p,t).
$$
Using Lemma \ref{npt_lemma} and the bound (\ref{eq_CT_R}), and recalling that $p = r/n$, we have, for $d \geq 3$, and for appropriate constants $c$ and $c'$, depending on $d$,
$$
S \leq \sum_{t=1}^{\infty}t^{d}\cdot c 2^{-t}n_{\sigma}(p/t,1) \leq c' \sum_{t=1}^{\infty}t^{d}2^{-t}\left(\frac{r}{t}\right)^{d-1}\log \left(\frac{r}{t}\right) \leq
$$
$$
c' \left(\sum_{t=1}^{\infty}t2^{-t}\right)r^{d-1}\log r = O(r^{d-1}\log r).
$$
For $d = 2$, arguing similarly, we have $S = O(r\alpha(r))$. This completes the proof of Theorem \ref{thm_cutting}. $\Box$


\section{Partition theorem for shallow points in $\reals^d$} \label{section_partition}

Let $P$ be a set of $n$ points in $\reals^d$. A \emph{simplicial partition} for $P$ is a collection $$\Pi = \biggl\{(P_{1},\Delta_{1}), \ldots,(P_{m},\Delta_{m})\biggr\}, $$
where the $P_{i}$'s are pairwise disjoint subsets (called the \emph{classes} of $\Pi$) forming a partition of $P$, and each $\Delta_{i}$ is a relatively open simplex containing $P_{i}$. We also require that the $P_{i}$'s be roughly of the same size, that is, $k \leq |P_{i}| \leq 2k$ for each $i$, for some parameter $k < n$ (so $m = \Theta(n/k)$). Assuming that the points of $P$ are in general position, and $k \geq d+1$, the simplices $\Delta_{i}$ are full-dimensional. We also note that the simplices $\Delta_{i}$ are not required to be pairwise disjoint, so a simplex $\Delta_{i}$ may also contain other points of $P$, in addition to those of $P_{i}$. (See however the recent work of Chan \cite{ch12} where the simplices can be made pairwise disjoint.)

For a hyperplane $h$, and a simplex $\Delta$, we say that $h$ \emph{crosses} $\Delta$ if $h \cap \Delta \neq \emptyset$ and $\Delta \not \subset h$ (thus a hyperplane does not cross a lower-dimensional simplex contained in it). We define the \emph{crossing number} of a hyperplane $h$ (with respect to $\Pi$) as the number of simplices $\Delta_{i}$ crossed by $h$, and define the \emph{crossing number} of $\Pi$ as the maximum crossing number of any hyperplane with respect to $\Pi$.


We recall two previous versions of the partition theorems, both developed by Matou\v{s}ek. The first one, used for obtaining an efficient simplex range counting algorithm, is the standard, general partition method. The second one, used for improved range reporting algorithms, is a partition method for shallow hyperplanes only, that is, the improved crossing number (relative to the first result), is guaranteed only for hyperplanes that contain up to a prescribed number of points on one side.

\begin{theorem} \label{thm_matousek_general}
\emph{{\bf(Partition Theorem \cite{mat92a})}}.
Let $P$ be an $n$-point set in $\reals^{d}$ and let $1 < k < n$ be a parameter. There exists a simplicial partition $\Pi$ for $P$ of size $O(n/k)$, whose classes satisfy $k \leq |P_{i}| \leq 2k$, with crossing number $O((n/k)^{1-1/d})$.
\end{theorem}

\begin{theorem} \label{thm_matousek_shallow}
\emph{{\bf(Partition Theorem for shallow hyperplanes \cite{mat92b})}}.
Let $P$ be an $n$-point set in $\reals^{d}$ and let $1< k < n$ be a parameter. There exists a simplicial partition $\Pi$ for $P$ of size $O(n/k)$, whose classes satisfy $k \leq |P_{i}| \leq 2k$, and such that the crossing number of any $k$-shallow hyperplane with respect to $\Pi$ is $O((n/k)^{1-1/ \lfloor d/2 \rfloor})$ for $d \geq 4$, and $O(\log (n/k))$ for $d = 2,3$.
\end{theorem}


Here is our version of the Partition Theorem, which caters to sets of shallow points and for any hyperplane.

\begin{theorem} \label{thm_partition}
\emph{{\bf(Partition Theorem for shallow points)}}.
Let $P$ be a set of $n$ $k$-shallow points in $\reals^d$, where $d \geq 2$ and $2\leq k \leq n/2$. There exists a simplicial partition $\Pi$ for $P$, whose classes $P_{i}$ satisfy $k \leq \left| P_{i} \right| \leq 2k$ $($so their number $m$ is $\Theta(n/k))$, and such that the crossing number of any hyperplane with respect to $\Pi$ is $O\left((n/k)^{1-1/(d-1)} \log^{2/(d-1)}(n/k)\right)$ for $d \geq 3$, and $O\left(\alpha(n/k)\log^{2}(n/k)\right)$ for $d = 2$.

Such a simplicial partition can be constructed in time $O(n^{1+\delta})$, for any fixed $\delta > 0$ $($where the constants in this bound and in the bound on the crossing number depend
on $\delta)$. The construction time can be improved to $O(n \log (n/k))$, provided that $k \geq n^{\gamma}$ for any fixed $\gamma > 0$ $($again with the constants in this bound and in the bound on the crossing number depending on $\gamma)$.
\end{theorem}

\noindent{\bf Remarks: (i)} Note that, in contrast with Theorem \ref{thm_matousek_general} and \ref{thm_matousek_shallow}, which hold for any $1 < k < n$, Theorem \ref{thm_partition} only holds for the shallowness parameter $k$ of the set $P$.
\noindent{\bf (ii)} We may assume that $n/k$ is larger than some suitable constant, since otherwise we can partition $P$ into a constant number ($\leq \lfloor n/k \rfloor$) of arbitrary classes, and the theorem then holds trivially.


\paragraph{}
The proof of Theorem \ref{thm_partition} is based on the following lemma, adapted from similar lemmas in \cite{mat92a,mat92b}.

\begin{lemma} \label{lemma_partition_Q}
Let $P$, $n$ and $k$ be as above, and let $Q$ be a given set of hyperplanes. Then there exists a simplicial partition $\Pi$ for $P$, each of whose classes $P_{i}$ satisfies $k \leq \left| P_{i} \right| \leq 2k$, such that the crossing number of every hyperplane $h \in Q$ with respect to $\Pi$ is
\begin{align*}
&O\left((n/k)^{1-1/(d-1)} \log^{2/(d-1)}(n/k) + \log |Q|\right), \quad \mbox{{\em for} $d \geq 3$, {\em and}} \\
&O\left(\alpha(n/k)\log^{2}(n/k) + \log |Q|\right), \quad \mbox{\emph{for} $d = 2$}.
\end{align*}
\end{lemma}


\noindent{\bf Proof.} We first present the proof for the case $d \geq 3$, and then consider the case $d = 2$, whose simpler proof uses the same machinery. We will inductively construct pairwise disjoint subsets $P_{1}, P_{2}, \ldots $ of $P$, and simplices $\Delta_{1}, \Delta_{2}, \ldots$, so that $P_{j} \subseteq \Delta_{j}$ for each $j$. Suppose that $P_{1}, \ldots, P_{i}$ have already been constructed, and set $P_{i}' = P\setminus(P_{1} \cup \cdots \cup P_{i})$ and $n_{i} = |P_{i}'|$, where we start with $P_{0}' = P$. We iterate the construction as long as $n_{i} > Bk$, for an appropriate constant $B$ that will be set below. Let $Z_{i}$ be a random sample of $P_{i}'$, where each point is chosen independently with probability $1/k$, and let $Y_{i}$ be the set of vertices of $Conv(Z_{i})$, as in Section \ref{section_sampling}. By Proposition \ref{lemma_prob}, with constant probability, $Y_{i}$ and $Z_{i}$ satisfy $|Y_{i}| \geq \frac{n_{i}}{8k}$ and $\frac{n_{i}}{2k} \leq |Z_{i}| \leq \frac{2n_{i}}{k}$ (here we use the assumption that $n_{i}/k$ is sufficiently large; see below). Assume that the sample $Z_{i}$ does satisfy these properties. Then, by Proposition \ref{prop_peps}, with probability at least $1 - q$, $Z_{i}$ is also a relative $(\frac{c_{2}k}{n_{i}} \log \frac{n_{i}}{c_{2}k}, \frac{1}{2})$-approximation for $(P_{i}',\R)$, where $\R$ is, as above, the set of all simplices in $\reals^{d}$, $c_{2} \geq 8 c_{1} (\delta + 1/q)$, $c_{1}$ is taken from Theorem \ref{thm_relative_peps}, and $\delta = O(d^{2})$ is the VC-dimension of $(P,\R)$. Let us set $c_{2} = \max \{(d + 1)/k, 8 c_{1} (\delta + 1/q)\}$, and assume that $c_{2} \geq 2$.

We distinguish between two cases. First suppose that $n_{i} > Bk$. For a hyperplane $h \in Q$, let $\kappa_{i}(h)$ denote the number of simplices among $\Delta_{1}, \ldots, \Delta_{i}$ crossed by $h$. We define a weighted collection $(Q, w_{i})$ by setting $w_{i}(h) = 2^{\kappa_{i}(h)}$ for each $h \in Q$.

We use a variant of the Cutting Lemma for weighted collections of hyperplanes. We fix a parameter $t_{i}$ (again, its value will be set shortly), and construct a $(1/t_{i})$-cutting $\Xi_{i}$ for the zone of the boundary $\sigma$ of the convex hull of $Y_{i}$ in the arrangement $\mathcal{A}(Q,w_{i})$. By Corollary \ref{corollary_weight_cutting}, $\Xi_{i}$ consists of at most $c_{3}t_{i}^{d-1} \log t_{i}$ simplices, for a suitable constant $c_{3}$. We set
$$
t_{i} = c_{4}\left(\frac{n_{i}/k}{\log^{2}(n_{i}/k)}\right)^{1/(d-1)},
$$
for an appropriate constant parameter $c_{4} \leq 1$, whose concrete value will be set later. Let $F$ be the solution of $F/\log^{2} F = (1/c_{4})^{d-1}$. We put $B = \max\{F, 2c_{2}, 256\}$ and note that (i) $t_{i} > 1$, (ii) $n_{i} \geq 256 k$ (as required by Proposition \ref{lemma_prob}), and (iii) $n_{i} \geq 2 c_{2} k$ (as required by Proposition \ref{prop_peps}). We estimate the size of $\Xi_{i}$ by
$$
|\Xi_{i}| \leq c_{3}t_{i}^{d-1}\log t_{i} \leq
c_{3}c_{4}^{d-1} \cdot \frac{n_{i}}{k} \cdot \frac{(d-1)\log c_{4} + \log\frac{n_{i}}{k} - 2 \log \log \frac{n_{i}}{k}}{(d-1) \log^{2}\frac{n_{i}}{k}} \leq  \frac{c_{3}c_{4}^{d-1}}{d-1}\cdot \frac{n_{i}}{k} \cdot \frac{1}{\log \frac{n_{i}}{k}}.
$$
(The final inequality holds when $\log c_{4} \leq 0$ and $\log \log (n_{i}/k) > 0$, which is indeed the case since $c_{4} \leq 1$ and $n_{i} > 4k$.) Using this bound, we argue that there exists a (relatively open) simplex $\tau \in \Xi_{i}$ containing at least $k$ points of $P_{i}'$. To see this, we first observe that each point of $Y_{i}$ is contained in some simplex of $\Xi_{i}$. Hence there exists a relatively open simplex $\tau \in \Xi_{i}$, such that $|Y_{i} \cap \tau| \geq |Y_{i}|/|\Xi_{i}|$. Since $|Y_{i}| \geq n_{i}/8k$, the measure of $\tau$ in $Z_{i}$ satisfies
$$
\overline{Z_{i}}(\tau) = \frac{|Z_{i} \cap \tau|}{|Z_{i}|} \geq \frac{|Y_{i} \cap \tau|}{|Z_{i}|} \geq {\frac{|Y_{i}|}{|\Xi_{i}|}} \cdot \frac{1}{|Z_{i}|} \geq
\frac{n_{i}}{8k} \cdot \frac{k}{2n_{i}} \cdot \frac{1}{|\Xi_{i}|} \geq
\frac{k}{n_{i}} \cdot \frac{(d-1)}{16 c_{3}c_{4}^{d-1}} \cdot \log \frac{n_{i}}{k} = A \frac{k}{n_{i}},
$$
where
\begin{equation} \label{eq_Z_measure}
A = \frac{d-1}{16 c_{3}c_{4}^{d-1}} \log \frac{n_{i}}{k}.
\end{equation}
We can now bound from below the number of points of $P_{i}'$ in $\tau$. Put $D = \max\{2,(d+1)/k\}$. Since $Z_{i}$ has the relative $\peps$-approximation property with respect to $P_{i}'$, with the values of $p$ and $\eps = 1/2$ as specified in Proposition \ref{prop_peps}, we have, by Definition \ref{definition_peps},
\begin{list} {(\roman{itemcounter})}{\usecounter{itemcounter} \leftmargin=2em}
\item If $\overline{P_{i}'}(\tau) \geq p = \frac{c_{2}k}{n_{i}} \log \frac{n_{i}}{c_{2}k}$, then, by the choice of $c_{2}$,
$$
|P_{i}' \cap \tau | \geq p n_{i} = c_{2}k \log \frac{n_{i}}{c_{2}k} \geq c_{2}k \geq \max \{d+1,2k\}.
$$
\item If $\overline{P_{i}'}(\tau) \leq p$, then
$$
\overline{P_{i}'}(\tau) \geq \overline{Z_{i}}(\tau) - \eps p \geq
\frac{k}{n_{i}} \left(A - \frac{c_{2}}{2}\log \frac{n_{i}}{c_{2}k}\right),
$$
\end{list}
and thus we have
$$
|P_{i}' \cap \tau | \geq k \left(A - \frac{c_{2}}{2}\log \frac{n_{i}}{c_{2}k}\right).
$$
By requiring that $A - \frac{c_{2}}{2}\log \frac{n_{i}}{c_{2}k} \geq D$, we get that $|P_{i}' \cap \tau| \geq \max\{d+1,2k\}$. Substituting the value of $A$ in (\ref{eq_Z_measure}), this requirement becomes
$$
\frac{d-1}{16 c_{3}c_{4}^{d-1}} \log \frac{n_{i}}{k} \geq D + \frac{c_{2}}{2}\log \frac{n_{i}}{c_{2}k},
$$
or
$$
\left(\frac{d-1}{16 c_{3}c_{4}^{d-1}} - \frac{c_{2}}{2}\right) \log \frac{n_{i}}{k} \geq D - \frac{c_2}{2}\log c_{2}.
$$
Since we assume that $n_{i} \geq 256 k$, this inequality will be satisfied if
\begin{equation} \label{eq_partition_constants}
\frac{d-1}{2 c_{3}c_{4}^{d-1}} \geq 4c_{2} + D - \frac{c_{2}}{2}\log c_{2}.
\end{equation}
This inequality in turn can be enforced by choosing $c_{4}$ to be a sufficiently small constant which depends only in $d$ (note that the dependence of $D$ on $k$ does not prevent us from choosing $c_{4}$ to be a constant).

\begin{figure}[htb]
    \begin{center}
        \scalebox{0.65}{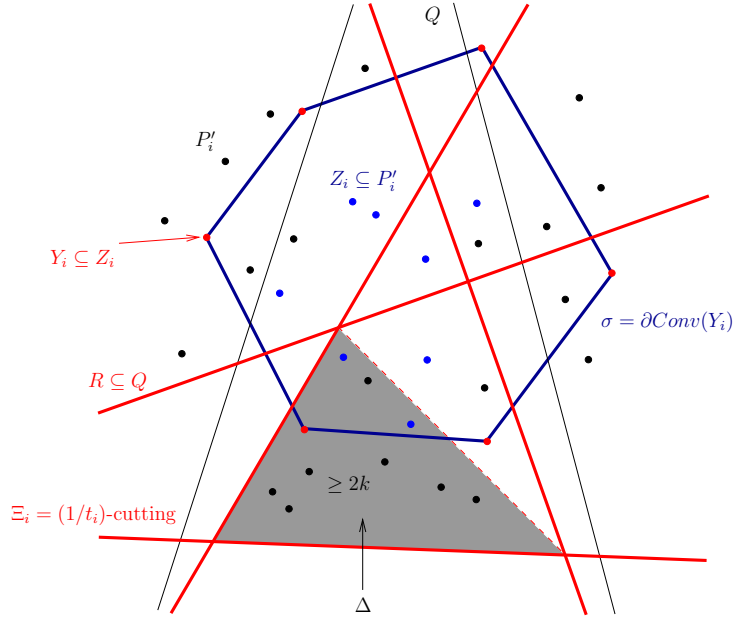}
        \caption{$\Delta$ is a simplex of the $(1/t_{i})$-cutting $\Xi_{i}$ of $(Q,w_{i})$ which contains $\geq 2k$ points of $P_{i}'$.}
        \label{fig_partition1}
    \end{center}
\end{figure}

To recap, with this choice of $c_{4}$ (and thus of $t_{i}$), we can ensure, with probability at least $1 - q$, that $|P_{i}' \cap \tau| \geq \max\{d+1,2k\}$, that is, $\tau$ is a full-dimensional simplex and contains at least $2k$ points of $P_{i}'$. We now set $P_{i+1}$ to be an arbitrary $k$-point subset of $P_{i}' \cap \tau$, and put $\Delta_{i+1} = \tau$, so $P_{i+1} \subseteq \Delta_{i+1}$. See Figure \ref{fig_partition1} for an illustration.

Proceeding with this construction, we reach an index $i = q$, for which $n_{q} \leq Bk$. We then partition the set $P_{q}'$ of the remaining points into at most $B$ arbitrary subsets $P_{q+1}, P_{q+2}, \ldots, P_{m}$ of size $k$ each, except for $P_{m}$, whose size is between $k$ and $2k - 1$. We define the simplices $\Delta_{q+1} = \Delta_{q+2} = \cdots = \Delta_{m} = \reals^{d}$, and set $\Pi = \left\{(P_{1}, \Delta_{1}), \ldots, (P_{m}, \Delta_{m})\right\}$, thereby completing the construction. Since the last phase adds at most a constant number of (i.e., at most $\lfloor B \rfloor$) classes and simplices, it only increases the crossing number of any hyperplane by an additive constant. Hence it suffices to bound the crossing number of any hyperplane relative to $\Pi' = \left\{(P_{1}, \Delta_{1}), \ldots, (P_{q}, \Delta_{q})\right\}$.


To do so, we estimate the final total weight $w_{q}(Q)$ of the hyperplanes of $Q$ in two different ways.

On one hand, the weight $w_{q}(h)$ of a hyperplane $h \in Q$ with crossing number $\kappa$ is equal to $2^{\kappa}$. Therefore, $\kappa  \leq \log w_{q}(h) \leq \log w_{q}(Q)$.

On the other hand, let us consider how $w_{i+1}(Q)$ increases compared to $w_{i}(Q)$. Note that, since $\Xi_{i}$ is a $(1/t_{i})$-cutting of the weighted collection $(Q,w_{i})$, it follows that $\Delta_{i+1}$ is crossed by a collection of hyperplanes, denoted by $Q^{*}$, whose total weight at step $i$ is at most $w_{i}(Q)/t_{i}$. When passing to step $i+1$, the weight of every hyperplane of $Q^{*}$ is doubled, whereas the weight of any other hyperplane remains unchanged. Thus, the total weight of $Q$ increases by at most $w_{i}(Q)/t_{i}$. That is,
\begin{align*}
w_{i+1}(Q) &\leq w_{i}(Q) + \frac{w_{i}(Q)}{t_{i}} = w_{i}(Q)\left(1 + \frac{1}{t_{i}} \right) \\
&\leq w_{i}(Q)\left(1 + \frac{1}{c_{4}} \left(\frac{\log^{2}(n_{i}/k)}{n_{i}/k} \right)^{1/(d-1)} \right).
\end{align*}
Let us put $r = n/k$, and recall that $w_{0}(Q) = |Q|$, $n_{i} = n - ik$, and $q \leq r$. Hence
$$
w_{q}(Q) \leq |Q|\prod_{i=0}^{q-1}\left(1 + \frac{1}{t_{i}} \right) \leq |Q|\prod_{i=0}^{q-1} \left(1 + \frac{1}{c_{4}} \left(\frac{\log^{2}(r-i)}{r-i}\right)^{1/(d-1)} \right).
$$
Taking logarithms and using the inequality $\ln(1+x) \leq x$, we get
$$
\log w_{q}(Q) \leq \log |Q| + \frac{1}{c_{4}\ln 2} \sum_{i=0}^{q-1}\left(\frac{\log^{2}(r-i)}{r-i}\right)^{1/(d-1)}
$$
\begin{equation} \label{eq_partition_step}
\leq \log |Q| + \frac{1}{c_{4}^{'}} \sum_{j=1}^{r}\left(\frac{\log^{2} j}{j}\right)^{1/(d-1)},
\end{equation}
where $c_{4}^{'} = c_{4}\ln 2$. That is, we have
\begin{align*}
\kappa & \leq \log w_{q}(Q) \leq \log|Q| + \frac{1}{c_{4}^{'}}\log^{2/(d-1)}r \sum_{j=1}^{r}\left(\frac{1}{j}\right)^{1/(d-1)} \\
& = O\left(\log|Q| + r^{1-1/(d-1)} \log^{2/(d-1)}r\right) \\
& = O \left(\log |Q| + (n/k)^{1 - 1/(d-1)} \log^{2/(d-1)}(n/k)\right),
\end{align*}
for $d \geq 3$. This completes the proof for $d \geq 3$.

Consider next the case $d = 2$, and follow the same reasoning as in higher dimensions. At step $i$ of the construction, we put
$$
t_{i} = c_{4} \cdot \frac{n_{i}/k}{\alpha(n_{i}/k) \log (n_{i}/k)},
$$
and construct a $(1/t_{i})$-cutting $\Xi_{i}$ of the zone of $Conv(Y_{i})$ in $\mathcal{A}(Q,w_{i})$, where $w_{i}$ is defined as above. By Corollary \ref{corollary_weight_cutting} we have
$$
|\Xi_{i}| \leq c_{3} t_{i}\alpha(t_{i}) \leq c_{3}c_{4} \cdot \frac{n_{i}}{k} \cdot \frac{1}{\log\frac{n_{i}}{k}},
$$
for an appropriate constant $c_{3}$. Hence, (\ref{eq_Z_measure}) and (\ref{eq_partition_constants}), with $d = 2$, are valid for this case as well. Let $F$ be the solution of $F/(\alpha(F)\log(F)) = 1/c_{4}$, and put $B = \max\{F, 2c_{2}, 256\}$ as in higher dimensions. Using the relative $\peps$-approximation property of $Z_{i}$ (which we may assume to hold), we obtain that there exists an open cell $\tau$ of $\Xi_{i}$, for which $|P_{i}' \cap \tau| \geq 2k$, provided that $c_{4}$ is chosen to be a sufficiently small constant, satisfying inequality (\ref{eq_partition_constants}). We take $P_{i+1}$ to be an arbitrary subset of $k$ points of $P_{i}' \cap \tau$, and put $\Delta_{i+1} = \tau$. We repeat this step until $n_{i} \leq Bk$, and then complete the construction as in the higher-dimensional case.

The analysis of the crossing number proceeds as above, except for the different value of the parameters $t_{i}$. Plugging this value into the appropriate variant of inequality (\ref{eq_partition_step}), yields the following bound
$$
\kappa \leq \log w_{q}(Q) \leq \log |Q| + \frac{1}{c_{4}^{'}} \sum_{j=1}^{r}\frac{\alpha(j) \log j }{j} \leq \log |Q| + \frac{1}{c_{4}^{'}} \alpha(r) \log r \sum_{j=1}^{r} \frac{1}{j} =
$$
$$
O\left(\log |Q| + \alpha(r) \log^{2}r \right) = O\left(\log |Q| + \alpha(n/k) \log^{2}(n/k) \right),
$$
where $c_{4}^{'} = c_{4}\ln 2$, as above. This concludes the proof of the lemma. $\Box$


\paragraph{}
The next step in the proof of Theorem \ref{thm_partition} is to choose a small `test-set' $Q$ of hyperplanes, with the property that the crossing number of any hyperplane is at most proportional to the maximum crossing number of a hyperplane in $Q$.

\begin{lemma} \label{lemma_test-set}
\emph{{\bf(Test-set lemma)}}.
Let $P$, $n$, and $k$ be as above. Then there exists a set $Q$ of $O((n/k)^{d/(d-1)})$ hyperplanes, such that, for any simplicial partition $\Pi = \{(P_{1},\Delta_{1}), \ldots,(P_{m},\Delta_{m})\}$ satisfying $\left| P_{i} \right| \geq k$ for every $i$, the following holds: If $\kappa_{0}$ is the maximum crossing number of a hyperplanes of $Q$ with respect to $\Pi$, then the crossing number of any hyperplane with respect to $\Pi$ is at most
$$
(d + 1)\kappa_{0} + O\left((n/k)^{1-1/(d-1)} \right),
$$
\end{lemma}
for $d \geq 3$, and $(d + 1)\kappa_{0} + 1$ for $d = 2$.

\noindent{\bf Proof.} Let $H = \mathcal{D}(P)$ be the collection of hyperplanes dual to the points of $P$. Put $t = \Theta((n/k)^{1/(d-1)})$. Construct a $(1/t)$-cutting $\Xi$ of size $O(t^{d})$ for $\mathcal{A}(H)$, and let $Q$ be the set of all hyperplanes dual to the vertices of $\Xi$. Clearly $|Q| = O((n/k)^{d/(d-1)})$.

Fix a simplicial partition $\Pi = \{(P_{1},\Delta_{1}), \ldots, (P_{m},\Delta_{m})\}$ as above.
Let $h$ be any hyperplane in the primal space. The point $\mathcal{D}(h)$ dual to $h$ is contained in a simplex $\sigma \in \Xi$. Let $G$ be the set of hyperplanes in the primal space dual to the vertices of $\sigma$. Clearly $G \subseteq Q$, so each of the $d+1$ hyperplanes of $G$ crosses at most $\kappa_{0}$ simplices of $\Pi$.

It remains to bound the number of simplices of $\Pi$ which are crossed by the hyperplane $h$ but by no hyperplane of $G$. Such simplices $\Delta_{i}$ must be completely contained in the zone of $h$ in the arrangement $\mathcal{A}(G)$, and hence this zone must also contain the points of their corresponding classes $P_{i}$ in its interior. It is elementary to verify (also see \cite{mat92a} and \cite{mat92b}), that any point of $P$ lying in the interior of the zone of $h$ in $\mathcal{A}(G)$ dualizes to a hyperplane of $H$ intersecting the interior of the simplex $\sigma$, and there are at most $n/t$ such hyperplanes in $H$. Hence, the zone of $h$ may contain at most these many points of $P$, implying that there are at most $n/(tk) = O((n/k)^{1-1/(d-1)})$ simplices of $\Pi$ completely contained in the zone of $h$ in $\mathcal{A}(G)$, since $\left| P_{i} \right| \geq k$ for every $i$. In the plane, there is at most $n/(tk) = 1$ such simplex. $\Box$


\noindent{\bf Proof of Theorem \ref{thm_partition}.} The proof of the Partition Theorem now follows easily. Consider first the case $d \geq 3$. Given the set $P$ of $n$ $k$-shallow points, we first construct a test-set $Q$ of $O((n/k)^{d/(d-1)})$ hyperplanes, as in Lemma \ref{lemma_test-set}. Then we apply Lemma \ref{lemma_partition_Q}, obtaining a simplicial partition $\Pi$ such that any hyperplane $h \in Q$ has crossing number
$$
O\left(\log |Q| + (n/k)^{1 - 1/(d-1)} \log^{2/(d-1)}(n/k)\right) = O\left((n/k)^{1 - 1/(d-1)} \log^{2/(d-1)}(n/k)\right)
$$
relative to $\Pi$. (Here we use the face that, when $n/k$ is at least some sufficiently large constant, $\log|Q|$ is dominated by the other term in the bound.) By the `test-set' property of $Q$, the crossing number of any hyperplane $h$ relative to $\Pi$ is at most
$$
(d+1) \cdot O\left((n/k)^{1 - 1/(d-1)} \log^{2/(d-1)}(n/k)\right) + O\left((n/k)^{1-1/(d-1)} \right)
$$
$$
= O\left((n/k)^{1-1/(d-1)} \log^{2/(d-1)}(n/k)\right),
$$
as claimed. The case $d = 2$ is argued similarly and the resulting crossing numbers, of the lines of $Q$, and thus also of any line, are $O(\alpha(n/k) \log^{2}(n/k))$. This completes the proof of the existence and main properties of the partition.


It remains to analyze the preprocessing time of this partition. We actually carry out the standard algorithm of \cite{mat92a}, which constructs a $(1/t_{i})$-cutting of the entire arrangement of $\mathcal{A}(Q,w_{i})$, and use the preceding analysis to argue that one of the simplices in the cutting will contain (with high probability) $2k$ points of $P_{i}'$. There is no need to explicitly construct $Z$, $Y$, and the zone of the convex hull of $Z$, they are needed only to guarantee the existence of such a simplex. With this approach, we can thus conclude that the preprocessing time of our simplicial partition for a set of $k$-shallow points is at most $O(n^{1+\delta})$ for any value of $k$ and for any fixed $\delta > 0$ (where the constants in this bound and in the bound on the crossing number both depend on $\delta$). If $k \geq n^{\alpha}$, for any fixed $\alpha > 0$, the preprocessing time can be reduced to $O(n \log (n/k))$, as in \cite{mat92a} (where the constants in both bounds now depend on $\alpha$). We note that, in the output-sensitive range counting algorithm below, we use the latter bound, since we deal there only with large values of $k$.

This concludes the proof of the Partition Theorem. $\Box$


Let us note the following special case of Theorem \ref{thm_partition}, where the points of $P$ are $1$-shallow.

\begin{corollary} \label{corollary_partition_1_shallow}
Let $P$ be a set of $n$ $1$-shallow points in $\reals^{d}$, where $d \geq 2$. There exists a simplicial partition $\Pi$ for $P$, whose classes $P_{i}$ satisfy $|P_{i}| = 2$ (except, possibly, for one class of size $3$), such that the crossing number of any hyperplane with respect to $\Pi$ is $O(n^{1-1/(d-1)} \log^{2/(d-1)}n)$ for $d \geq 3$, and $O(\alpha(n)\log^{2}n)$ for $d = 2$.
\end{corollary}


\section{Applications} \label{section_applications}

\paragraph{Overview.}In this section we derive two major applications of the Partition Theorem for shallow sets (Theorem \ref{thm_partition}) presented in the previous section. The first application is the construction of a spanning tree with small relative crossing number, which depends on the shallowness of the crossing hyperplane. We then turn this construction into a construction of a relative $\peps$-approximation, using the same machinery of Har-Peled and Sharir \cite{hs11}, thus extending their planar construction to higher dimensions. The second application is an output-sensitive halfspace range counting data structure, where the query time is better than that of the standard algorithm of Matou\v{s}ek \cite{mat92a}, when the hyperplane bounding the query halfspace is shallow.


\subsection{Spanning trees with small relative crossing number in $\reals^d$} \label{subsection_spanning_tree}

In our first application, we extend the ``weight-sensitive'' version of \emph{spanning trees with small relative crossing number} in the plane, studied by Har-Peled and Sharir \cite{hs11}, to higher dimensions. Both our extension and the construct in \cite{hs11} refine the following classical construct of spanning trees with small crossing number, as obtained by Chazelle and Welzl \cite{cw89}, with a simplified construction given later in \cite{wel92}.

\begin{theorem} \emph{{\bf\cite{cw89, wel92}}} \label{thm_standard_spanning_T}
Let $P$ be a set of $n$ points in $\reals^{d}$, $d \geq 2$. Then there exists a straight-edge spanning tree $\Tree$ of $P$ such that each hyperplane in $\reals^{d}$ crosses at most $O(n^{1-1/d})$ edges of $\Tree$.
\end{theorem}


We begin by refining this result for a set of $k$-shallow points.

\begin{lemma} \label{lemma_spanning_T_k}
Let $P$ be a set of $n$ $k$-shallow points in $\reals^{d}$. One can construct a spanning tree $\EuScript{T}$ for $P$, such that any hyperplane crosses at most $O\left(n^{1-1/(d-1)}k^{1/d(d-1)}\log^{2/(d-1)}(n/k)\right)$ edges of $\Tree$, for $d \geq 3$, and  $O\left(\alpha(n/k)\log^{2}(n/k)\left(\sqrt{k} + \log(n/k)\right)\right)$ edges, for $d = 2$.
\end{lemma}

\noindent{\bf Proof.} We first consider the case $d \geq 3$. Given a set $P$ of $n$ $k$-shallow points, we construct a simplicial partition $\Pi = \{(P_{1},\Delta_{1}), \ldots,(P_{m},\Delta_{m})\}$ of $P$, by applying the Partition Theorem (Theorem \ref{thm_partition}), for sets of $2k$-shallow points (clearly, the points of $P$ are also $(2k)$-shallow). (We note that replacing $k$ by $2k$ does not affect asymptotically the crossing number of $\Pi$.) Each class $P_{i}$ of the partition now satisfies $2k \leq P_{i} \leq 4k$.

Given $\Pi$, we first ignore its ``tail'', namely the pairs $(P_{q+1},\Delta_{q+1}), \ldots, (P_{m}, \Delta_{m})$, where $q$, as in the proof of Lemma \ref{lemma_partition_Q}, is the first index for which $n_{q} < B\cdot(2k) = B'k$. Let $\Pi' \subseteq \Pi$ denote the collection $\{(P_{1},\Delta_{1}),
\ldots, (P_{q},\Delta_{q})\}$. For each $i = 1, \ldots, q$, we construct a spanning tree $\Tree_{i}$ for $P_{i}$ with crossing number $O(k^{1-1/d})$ using Theorem  \ref{thm_standard_spanning_T}. Then we connect those trees by segments, called \emph{bridges}, into a single tree, whose construction will be detailed below. Finally, we construct another spanning tree, denoted by $\Tree_{q+1}$, of the at most $B'k$ points in $P_{q+1}
\cup \cdots \cup P_{m}$, with crossing number $O((B'k)^{1-1/d}) = O(k^{1-1/d})$. We connect $\Tree_{q+1}$ by a single edge to an arbitrary point in the other tree. The union of the trees $\Tree_{1}, \ldots, \Tree_{q+1}$ together with the connecting bridges forms our spanning tree $\Tree$.

The connecting bridges are constructed as follows. We pick a point $p_{i}$ from $P_{i}$, uniformly at random, for each $i=1,\ldots,q$. Let $R$ denote the resulting set of $m = \Theta(n/k)$ points. The probability of any single point to belong to $R$ is at most $1/(2k)$. (It is here that we use the larger size of the classes; see below.) Suppose that some point $a \in P$ has been chosen in $R$. By assumption, $a$ is $k$-shallow, so there is a halfspace $H$ which contains $a$ and at most $k-1$ other points of $P$. Since each of these points appears in $R$ with probability $\leq 1/(2k)$, the expected number of points in $R \cap H$ (conditioned on $a$ being chosen in $R$) is $< 1/2$. Hence, using Markov's inequality, with (conditional) probability at least $1/2$, there is no other point apart from $a$ in $R \cap H$, that is, $a$ is a vertex of the convex hull of $R$. Hence, the expected number of vertices of $Conv(R)$ is at least $|R|/2$. We may assume that $R$ satisfies this property, and denote by $R_{0}$ the subset of hull vertices of $Conv(R)$; Thus, $|R_{0}| \geq m/2$.

Apply Corollary \ref{corollary_partition_1_shallow} to the set $R_{0}$, to obtain a partition of it into $\frac{1}{2}|R_{0}| \geq \frac{m}{4} - 1$ disjoint pairs. Let $E$ denote the collection of the segments connecting the points in each pair. By construction, each hyperplane crosses at most  $O(|R_{0}|^{1-1/(d-1)} \log^{2/(d-1)}|R_{0}|)$ segments of $E$.

The segments of $E$ merge together some subsets of trees $\Tree_{i}$ into larger trees. Since we created at least $\frac{m}{4} - 1$ bridges, the number of disconnected trees is now $m' \leq m - ((m/4) - 1) = 3m/4 + 1$. We now repeat the above construction to the new trees, or, rather, to the subsets of $P$ that they span. That is, we choose a random point from each subset, take the subset of hull vertices of the resulting sample, and apply Corollary \ref{corollary_partition_1_shallow} to it, to create new bridges. The situation has actually improved, because these vertex sets of the current collection of trees are now larger (while the points are still $k$-shallow), so the probability of choosing any specific point is smaller; hence the probability of a sampled point to become a hull vertex can only grow. In any case, we create at least $\frac{m'}{4} - 1$ new bridges, and keep iterating in this manner until all trees have been merged into a single tree.


As is easily checked, the number of bridges crossed by a hyperplane $h$ is at most
$$
O\left(\sum_{j \geq 0} \left(\left(\tfrac{3}{4}\right)^{j} \tfrac{n}{k} \right)^{1-1/(d-1)} \log^{2/(d-1)} \left(\left(\tfrac{3}{4} \right)^{j} \tfrac{n}{k} \right)\right)
= O\left((n/k)^{1-1/(d-1)} \log^{2/(d-1)}(n/k)\right).
$$
In addition, $h$ crosses $O\left((n/k)^{1-1/(d-1)} \log^{2/(d-1)}(n/k)\right)$ of the simplices $\Delta_{i}$, and, within each such simplex $\Delta_{i}$, it crosses $O(k^{1-1/d})$ edges of the corresponding tree $\Tree_{i}$. Also, it crosses $O(k^{1-1/d})$ edges of the tree $\Tree_{q+1}$. Since the bound within the simplices dominates the other two bounds, we conclude that $h$ crosses at most
$$
O\left((n/k)^{1-1/(d-1)} \log^{2/(d-1)}(n/k) \cdot k^{1-1/d}\right) = O\left(n^{1-1/(d-1)} k^{1/d(d-1)} \log^{2/(d-1)}(n/k)\right)
$$
edges of $\Tree$. This completes the proof for $d \geq 3$.

For $d = 2$, follow the same construction as described above, and re-estimate the crossing number of a hyperplane $h$ relative to our tree $\Tree$. We observe that $h$ crosses edges from $O\left(\alpha(n/k) \log^{2}(n/k)\right)$ trees of $\Tree_{i}$, and within each of them it crosses $O(\sqrt{k})$ edges. In addition, $h$ crosses $O(\sqrt{k})$ edges of $\Tree_{q+1}$. Following the same mechanism of the bridge construction and its analysis, the number of bridges crossed by $h$ is at most
\begin{align*}
O\left(\sum_{j \geq 0}\alpha \left(\left(\tfrac{3}{4}\right)^{j} \tfrac{n}{k} \right)
\log^{2}\left(\left(\tfrac{3}{4}\right)^{j}\tfrac{n}{k}\right) \right)
& = O\left(\alpha(n/k) \sum_{j \geq 0}\left(\log(n/k) - j \right)^{2}\right) \\
& = O\left(\alpha(n/k) \log^{3}(n/k) \right).
\end{align*}
Thus, altogether, the crossing number of $\Tree$ is bounded by
$$
O\left(\alpha(n/k) \log^{2}(n/k) (\sqrt{k} + \log(n/k))\right),
$$
as asserted. $\Box$


\paragraph{}
Let now $P$ be a set of $n$ points in $\reals^{d}$ (without the shallowness assumption). For a (non vertical) hyperplane $h$, let $w_{h}^{+}$ (resp., $w_{h}^{-}$) be the number of points of $P$ lying above (resp., below or on) $h$, and define the \emph{weight} of $h$, denoted by $w_{h}$, to be $\min\{w_{h}^{+}, w_{h}^{-}\}$.

\begin{theorem} \label{thm_spanning_T}
Let $P$ be a set of $n$ points in $\reals^{d}$. Then one can construct a spanning tree $\EuScript{T}$ for $P$, such that any hyperplane $h$ crosses at most $O\left(n^{1-1/(d-1)}w_{h}^{1/d(d-1)}\log^{2/(d-1)}(n/w_{h})\right)$ edges of $\EuScript{T}$, for $d \geq 3$, or $O\left(\sqrt{w_{h}}\alpha(n/w_{h})\log^{2}(n/w_{h}) + \alpha(n)\log^{4}n \right)$ edges, for $d = 2$.
\end{theorem}

\noindent{\bf Proof.} We construct a sequence of subsets of $P$, as follows. Put $P_{0}' = P$. At the $i^{th}$ step, $i \geq 1$, let $P_{i}$ denote the set of (at most) $2^{i}$-shallow points of $P_{i-1}'$, and put $P_{i}' = P  \setminus (P_{1}\cup \cdots \cup P_{i})$. We stop when $P_{i}'$ becomes empty. By construction, the $i^{th}$ step removes at least $2^{i}$ points from $P_{i-1}'$, because any (exactly) $2^{i}$-shallow halfspace $H$ contains $2^{i}$ points of $P_{i}'$, all of which are (at most) $2^{i}$-shallow. Hence, $|P_{i}'| \leq |P_{i-1}'| - 2^{i}$, and so the process terminates after $O(\log n)$ steps. At the $i^{th}$ step, we construct a spanning tree $\Tree_{i}$ for $P_{i}$, using Lemma \ref{lemma_spanning_T_k}, with $k = 2^{i}$. Connect the resulting trees by $O(\log n)$ additional straight segments (in an arbitrary manner) into a single spanning tree $\Tree$ of $P$.

We claim that $\Tree$ is the desired spanning tree. Indeed, consider an arbitrary hyperplane $h$ of weight $w_{h} = k$. We observe that $h$ cannot cross any of the trees $\Tree_{i}$, for $i > U = \lceil \log_{2} k \rceil$. To see this, assume to the contrary that $h$ crosses $\Tree_{j}$ for some $j > U$. That is, there exist two points $p_{1}, p_{2} \in P_{j}$ which are separated by $h$. Without loss of generality, assume that $p_{1}$ lies in the halfspace bounded by $h$ which contains $k$ points. In particular, $p_{1}$ must be $k$-shallow in $P$ (and thus also in any subset $P_{i}'$ containing it), so it must have been removed at some step $i \leq U$ and thus cannot belong to $P_{j}$.

Thus, $h$ crosses only the first $U$ trees of our construction. Hence, for $d \geq 3$, the number of edges of $\Tree$ that $h$ crosses, excluding the $O(\log n)$ connecting edges, is at most
\begin{align*}
& \sum_{i=1}^{U} O\left(n^{1-1/(d-1)}(2^{i})^{1/d(d-1)} \log^{2/(d-1)}(n/2^{i})\right) \\
& = O\left(n^{1-1/(d-1)}\sum_{i=1}^{U}\left(2^{1/d(d-1)}\right)^{i}(\log n - i)^{2/(d-1)}\right) \\
& = O\left(n^{1-1/d}\sum_{j=\log n - U}^{\log n - 1}\frac{j^{2/(d-1)}}{\left(2^{1/d(d-1)}\right)^{j}}\right).
\end{align*}
We have $a = \log n - U = \log\frac{n}{k}-\delta$, for some $0 \leq \delta \leq 1$, and we can estimate the sum by the integral
$$
\int_{a}^{\infty} \frac{x^{2/(d-1)}}{(2^{1/d(d-1)})^{x}} dx.
$$
By integrating in parts, it easily follows that, for $a \geq 1$, $0 < u \leq 1$, and $w > 1$,
\begin{equation} \label{eq_integral}
\int_{a}^{\infty}\frac{x^{u}}{w^{x}}dx \leq \frac{1}{\ln w} \cdot \frac{a^{u}}{w^{a}} + \frac{u}{\ln^{2}w} \cdot \frac{1}{w^{a}} = O\left(\frac{a^{u}}{w^{a}}\right),
\end{equation}
where the $O(\cdot)$ notation is with respect to the growth of $a$. This allows us to bound the sum above by
$$
O\left(n^{1-1/d}(k/n)^{1/d(d-1)} \log^{2/(d-1)}(n/k) \right) = O\left(n^{1-1/(d-1)}k^{1/d(d-1)} \log^{2/(d-1)}(n/k) \right),
$$
which is the asserted bound on the crossing number of $h$, for $d \geq 3$.

Let us consider the case $d = 2$. The crossing number of a line $l$ relative to $\Tree$ is bounded by
$$
\sum_{i=1}^{U}O\left(\alpha(n/2^{i}) \log^{2}(n/2^{i}) \left(\sqrt{2^{i}} + \log(n/2^{i})\right) \right),
$$
which we separate into the two sums:
$$
I_{1} = \sum_{i=1}^{U} O\left(\sqrt{2^{i}} \alpha(n/2^{i}) \log^{2}(n/2^{i})\right), \quad \mbox{and} \quad I_{2} = \sum_{i=1}^{U} O\left(\alpha(n/2^{i})\log^{3}(n/2^{i})\right).
$$
$I_{1}$ is bounded by
\begin{equation} \label{eq_alpha_sum}
\sum_{i=1}^{U}O\left(\sqrt{2^{i}}(\log n - i)^{2}\alpha(n/2^{i})\right) =
O\left(\sqrt n \sum_{j=\log n - U}^{\log n - 1}\frac{j^{2}\alpha(2^{j})}{\sqrt{2^{j}}}\right).
\end{equation}
$\alpha(\cdot)$ is very slowly increasing; in particular, it satisfies $\alpha(2x) \leq \alpha(x)+1$ for any $x$, and so, for $\alpha(x) \geq 3$, we have
$$
\frac{\alpha(2x)}{\alpha(x)} \leq 1 + \frac{1}{\alpha(x)} \leq \frac{4}{3}.
$$
We can therefore bound the sum in (\ref{eq_alpha_sum}), denoted by $S$, by
$$
S \leq \alpha(n/k) \sum_{j=\log n - U}^{\log n - 1} \frac{j^{2} \left(\tfrac{4}{3}\right)^{j - \log n + U}}{\sqrt{2^{j}}} \leq \alpha(n/k) \left(\tfrac{3}{4}\right)^{\log n - U} \sum_{j=\log n - U}^{\log n - 1} j^{2} \left(\tfrac{4}{3\sqrt{2}}\right)^{j}.
$$
Using again the integral bound in (\ref{eq_integral}), we obtain, for an appropriate constant $c_{1}$,
$$
S \leq c_{1} \cdot \alpha(n/k) \left(\tfrac{3}{4}\right)^{\log n - U} \log^{2}(n/k) \left(\tfrac{4}{3\sqrt{2}}\right)^{\log n - U} = O\left(\sqrt{\frac{k}{n}} \alpha(n/k) \log^{2}(n/k)\right),
$$
from which we obtain
$$
I_{1} = O\left(\sqrt{k} \alpha(n/k) \log^{2}(n/k)\right).
$$
For $I_{2}$, we use the trivial bound
$$
I_{2} = \sum_{i=1}^{U} O\left(\alpha(n/2^{i})\log^{3}(n/2^{i})\right) = O\left(\alpha(n)\log^{4}n \right).
$$
Altogether, the crossing number of $l$ is at most
$$
O\left(\sqrt{k}\alpha(n/k)\log^{2}(n/k) + \alpha(n)\log^{4}n \right).
$$
This establishes the bound on the crossing number for $d = 2$. $\Box$

\noindent{\bf Remark:} Note that the bound in Theorem \ref{thm_spanning_T} for the planar case is slightly worse than the bound derived in \cite{hs11}. Note also that we have worked harder on the estimation of $I_{1}$ to ensure that when $k = \Theta(n)$ the bound coincides with the standard bound $O(\sqrt{n})$.


\subsubsection{Relative $\peps$-approximations for halfspaces}

We next turn the above construction of a spanning tree with small relative crossing number into a construction of a relative $\peps$-approximation for a set of points in $\reals^{d}$ and for halfspace ranges. We base our construction on the machinery of Har-Peled and Sharir \cite{hs11}, thus extending their planar construction to higher dimensions.

Let $P$ be a set of $n$ points in $\reals^{d}$, $d \geq 3$, and let $\Tree$ be a spanning tree of $P$ as provided in Theorem \ref{thm_spanning_T}. We replace $\Tree$ by a perfect matching $M$ of $P$, with the same asymptotic relative crossing number, i.e., the number of pairs of $M$ that are separated by a hyperplane of weight $k$ is at most $O\left(n^{1-1/(d-1)} k^{1/d(d-1)} \log^{2/(d-1)}(n/k)\right)$. This is done in a standard manner --- we first convert $\Tree$ to a spanning path whose relative crossing number by any hyperplane is at most twice the crossing number of the same hyperplane with $\Tree$, and then pick every other edge of the path. To simplify the presentation, and to ensure that the resulting collection of edges is indeed a perfect matching, we assume that $n$ is even.

We now construct a coloring of $P$ with low discrepancy, by randomly coloring the points in each pair of $M$. Specifically, each pair is randomly and independently colored either as $-1,+1$ or as $+1,-1$, with equal probabilities. The standard theory of discrepancy (see \cite{cha01} and \cite{mat99}) yields the following variant.

\begin{lemma} \label{lemma_discrepancy}
Given a set $P$ of $n$ points in $\reals^{d}$, $d \geq 3$, one can construct a coloring $\chi:\;P\mapsto \{-1,1\}$, such that, for any halfspace $H$,
$$
\chi(P \cap H) = \sum_{p \in P \cap H}\chi(p) = O\left(n^{(d-2)/2(d-1)} |P \cap H|^{1/2d(d-1)} \log^{(d+1)/2(d-1)}n \right).
$$
The coloring is balanced --- each color class consists of exactly $n/2$ points of $P$.
\end{lemma}

\noindent{\bf Proof.} Indeed (see \cite{cha01, mat99}), if the maximum crossing number is $X$ then the above discrepancy can be bounded by $O(\sqrt{X\log n})$, and the asserted bound is then immediate by Theorem \ref{thm_spanning_T}. $\Box$

We also need the following technical lemma, taken from \cite{hs11}.
\begin{lemma} \label{lemma_approx_technical}
For any $x \geq 0$, $y > 0$, and $0 < p < 1$, we have $x^{p} < (x+y)/y^{1-p}$.
\end{lemma}

As we next show, the improved discrepancy bound of Lemma \ref{lemma_discrepancy} leads to an improved bound of the size of $(\nu,\alpha)$-samples for our range space, and, consequently, for the size of relative $\peps$-approximations (with some constraints on the relationship between $\eps$ and $p$, as noted below). Let us introduce the following parameters:
$$
\gamma = \frac{2d(d-1)-1}{(d-1)(d+1)}, \quad \quad \mu = \frac{2d}{d+1}, \quad \text{and} \quad \quad \eta = \frac{d}{d-1}.
$$

\begin{theorem} \label{thm_improved_relative_peps}
Given a set $P$ of $n$ points in $\reals^{d}$, $d \geq 3$, and parameters $0 < p < 1$ and $0 < \eps < 1$, one can construct a relative $(p,\eps)$-approximation $Z \subseteq P$ of size
$$
O\left(\frac{1}{\eps^{\mu} p^{\gamma}} \log^{\eta} \frac{1}{\eps p}\right).
$$
\end{theorem}

We observe that, ignoring the power $\eta > 1$ of the logarithmic factor, this bound is an improvement of the bound in Theorem \ref{thm_relative_peps}, as long as $\eps$ and $p$ satisfy
$$
\frac{1}{\eps^{\mu}p^{\gamma}} < \frac{1}{\eps^{2}p}, \quad\quad \text{or} \quad\quad \eps < p^{\frac{\gamma-1}{2-\mu}}.
$$
Substituting the values of $\gamma$ and $\mu$, we have
$$
\eps < p^{\frac{d(d-2)}{2(d-1)}},
$$
which is the required constraint on the relationship between $\eps$ and $p$, for which Theorem \ref{thm_improved_relative_peps} does indeed yield an improvement (modulo the small "penalty" in the logarithmic factor) . We also note that for $d = 2$ the bound in the theorem coincides with the bound in \cite{hs11} (except for the power of the logarithmic factor).

\noindent{\bf Proof.} Following one of the classical constructions of $\eps$-approximations (see \cite{cha01}), we repeatedly halve $P$, until we obtain a subset of size as asserted in the theorem, and then argue that the resulting set has the desired approximation property. Formally, we set $P_{0} = P$, and, at the $i^{th}$ step, partition $P_{i-1}$ into two equal halves, using Lemma \ref{lemma_discrepancy}; let $P_{i}$ and $P_{i}'$ denote the two halves (consisting of the points that are colored $+1$, $-1$, respectively). We keep $P_{i}$, remove $P_{i}'$, and continue with the halving process. Let $n_{i} = |P|/2^{i}$ denote the size of $P_{i}$. For any halfspace $H$, we have
\begin{align*}
\left| |P_{i} \cap H| - |P_{i}' \cap H|\right| & \leq c_{1} \cdot n_{i-1}^{(d-2)/2(d-1)} \cdot |P_{i-1} \cap H|^{1/2d(d-1)} \cdot \log^{\frac{d+1}{2(d-1)}}n_{i-1} \\
& \leq c_{2} \cdot n_{i}^{(d-2)/2(d-1)} \cdot |P_{i-1} \cap H|^{1/2d(d-1)} \cdot \log^{\frac{d+1}{2(d-1)}}n_{i},
\end{align*}
for appropriate constants $c_{1}$ and $c_{2}$. Recalling out notation
$$
\meas{P_{i}}{H} = \frac{|P_{i} \cap H|}{|P_{i}|} \quad \text{and} \quad \meas{P_{i}'}{H} = \frac{|P_{i}' \cap H|}{|P_{i}'|},
$$
this can be rewritten as
$$
|\meas{P_{i}}{H} - \meas{P_{i}'}{H}| \leq c_{2} \frac{|P_{i-1} \cap H|^{1/2d(d-1)}}{n_{i}^{1 - (d-2)/2(d-1)}} \log^{\frac{d+1}{2(d-1)}} n_{i}= c_{3} \frac{ \meas{P_{i-1}}{H}^{1/2d(d-1)}}{n_{i}^{(d+1)/2d}} \log ^{\frac{d+1}{2(d-1)}}n_{i},
$$
for an appropriate constant $c_{3}$. Since $|P_{i}| = |P_{i}'|$ and $P_{i-1} = P_{i} \cup P_{i}'$, we have
$$
\meas{P_{i-1}}{H} = \frac{|P_{i-1} \cap H|}{|P_{i-1}|} = \frac{|P_{i} \cap H|}{2|P_{i}|} + \frac{|P_{i}' \cap H|}{2|P_{i}'|} = \frac{1}{2}\left(\meas{P_{i}}{H} + \meas{P_{i}'}{H}    \right).
$$
Combining together the last two relations, we have
$$
|\meas{P_{i-1}}{H} - \meas{P_{i}}{H}| = \frac{|\meas{P_{i}}{H} - \meas{P_{i}'}{H}|}{2} \leq
c_{4} \frac{\meas{P_{i-1}}{H}^{1/2d(d-1)}}{n_{i}^{(d+1)/2d}} \log^{\frac{d+1}{2(d-1)}}n_{i},
$$
where $c_{4} = c_{3}/2$. Applying Lemma \ref{lemma_approx_technical}, with $p = 1/2d(d-1)$, $x = \meas{P_{i-1}}{h}$, and $y = p$, the last expression is bounded by
$$
\frac{c_{4} \cdot \log^{\frac{d+1}{2(d-1)}}n_{i}}{n_{i}^{(d+1)/2d}} \cdot \frac{\meas{P_{i-1}}{H} + p }{p^{1-1/2d(d-1)}} \leq \frac{c_{4} \cdot \log^{\frac{d+1}{2(d-1)}}n_{i}}{n_{i}^{(d+1)/2d} p^{1-1/2d(d-1)}} \left(\meas{P_{i-1}}{H} + \meas{P_{i}}{H} + p \right).
$$
This implies that, in the notation of Section \ref{subsection_approximations},
$$
d_{p}(\meas{P_{i-1}}{H}, \meas{P_{i}}{H}) \leq \frac{c_{4} \cdot \log^{\frac{d+1}{2(d-1)}} n_{i}}{n_{i}^{(d+1)/2d} p^{1-1/2d(d-1)}}.
$$
Since $d_{p}$ is a metric, the triangle inequality implies that
\begin{align*}
d_{p}(\meas{P}{H}, \meas{P_{i}}{H}) &\leq \sum_{k=1}^{i}d_{p}(\meas{P_{k-1}}{H}, \meas{P_{k}}{H}) \leq \frac{c_{4}}{p^{1-1/2d(d-1)}}\sum_{k=1}^{i}\frac{\log^{\frac{d+1}{2(d-1)}} n_{k}}{n_{k}^{(d+1)/2d}} \\
&= O\left(\frac{\log^{\frac{d+1}{2(d-1)}}n_{i}}{p^{1-1/2d(d-1)} n_{i}^{(d+1)/2d}}\right).
\end{align*}
We can therefore ensure that $d_{p}(\meas{P}{H}, \meas{P_{i}}{H}) < \eps$, for every halfspace $H$, provided that
$n_{i} = \Omega\left(\frac{1}{\eps^{\mu} p^{\gamma}} \log^{\eta}\frac{1}{\eps p}\right)$. Taking $Z$ to be the smallest $P_{i}$ which still satisfies this size constraint, i.e., of size  $\Theta\left(\frac{1}{\eps^{\mu} p^{\gamma}} \log^{\eta}\frac{1}{\eps p}\right)$, we have thus shown that $Z$ is a $(p,\eps)$-sample. As follows from the equivalence between $(p,\eps)$-samples and relative $(p,\eps)$-approximations (see \cite{hs11} and Theorem \ref{thm_relative_peps}), $Z$ is also a relative $(p,\eps)$-approximation, which completes the proof. $\Box$

\noindent{\bf Remark.}
Very recent work by Ezra~\cite{ez14} obtains improved bounds for the
size of relative approximations in a more general setting, which also
includes the case of halfspace ranges in higher dimensions.
For this latter setup, the bound obtained in \cite{ez14} is
$$
\max \left\{ O\left(\log^{\frac{3d+1}{d+1}} n \right) , \;
O\left(\frac{ \log^{\frac{3d+1}{d+1}} \frac{1}{\eps p}}
{ p^{\frac{d+\lfloor d/2 \rfloor}{d+1}} \eps^{\frac{2d}{d+1}} }
\right) \right\} ,
$$
where the power of $1/p$ is smaller than the one that we have obtained.


\subsection{Output-sensitive halfspace range counting in $\reals^d$} \label{subsection_range_countning}

Our second main application is an improved, output-sensitive algorithm for halfspace range counting in $\reals^{d}$, for any $d \geq 3$. That is, we are given a set $P$ of $n$ points in $\reals^{d}$, and want to preprocess it into a data structure that can answer efficiently queries, in which we are given a halfspace $H$, and we wish to count $|P \cap H|$ (exactly). We will only consider the case where we seek a data structure with near-linear storage. Our goal is an algorithm whose query performance is sensitive to the output size $w_{H} = |P \cap H|$.

Before describing our solution, we recall that when the output size $w_{H}$ is reasonably small, we can trivially turn the halfspace range reporting algorithm of Matou\v{s}ek \cite{mat92b} into a range counting algorithm. This algorithm uses a data structure with $O(n \log \log n)$ storage, which can be constructed in $O(n \log n)$ time, and a halfspace range reporting (counting) query with a halfspace $H$ can be answered in time $O(n^{1-1/\lfloor d/2 \rfloor} \log^{O(1)}n + w_{H})$. \footnote{Note that, if we allow subtraction, we can achieve the same query cost when the query halfspace is the complementary one, which contains $n - w_{H}$ points.} However, when $w_{H}$ is large, this becomes quite inefficient. Our solution will use Matou\v{s}ek's algorithm as one of its components, but its novel contribution is for the case where $w_{H}$ is large; see below for details.

Here is an informal overview of the algorithm. We partition $P$ into a logarithmic number of subsets $P_{1}, \ldots , P_{l}$, so that, for each $i$, $P_{i}$ consists of points which are $O(k_{i})$-shallow in $P \setminus (P_{1} \cup \cdots \cup P_{i-1})$, where $k_{1}$ is specified below, and $k_{i+1} = 2k_{i}$, for $i \geq 1$. This partition, as we show later, has the property that a halfspace $H$ of weight $w_{H}$ will miss the convex hulls of all the sets $P_{i}$, for which $k_{i} \geq 2 w_{H}$. Handling sets $P_{i}$ with $k_{i} < 2w_{H}$ is done using the partition data structure for shallow sets, as provided in Theorem \ref{thm_partition}. This will result in a query time bound which depends on $w_{H}$ in roughly the same manner as the bounds in Theorem \ref{thm_spanning_T}. Specifically, we show:

\begin{theorem} \label{thm_range_counting}
Given a set $P$ of $n$ points in $\reals^{d}$, $d \geq 3$, one can construct a data structure for the halfspace range counting problem for $P$, so that, for any query halfspace $H$, the number $w_{H} = |P \cap H|$ of points of $P$ in $H$ can be counted, for $d \geq 4$, in time
\begin{equation} \label{eq_range_counting_time}
Q(n) =
\begin{cases}
    O\left(n^{1-\lfloor d/2 \rfloor} \log^{O(1)}n + w_{H}\right), & \mbox{for $w_{H} \leq c_{1} n^{1-(d-1)/(d(d-1)-1)}$}, \\
    O\left(n^{1-1/(d-1)} w_{H}^{1/d(d-1)} \log^{O(1)}n \right), & \mbox{otherwise},
\end{cases}
\end{equation}
for an appropriate constant $c_{1}$. For $d = 3$, the first bound for the query time is $O(\log n + w_{H})$, and the second bound remains the same.

The data structure uses $O(n\log\log n)$ storage, and preprocessing time
$$
T(n) =
\begin{cases}
    O\left(n^{1 + \frac{(d-2)(d-1)}{2(d(d-1)-1)}} \log^{O(1)}n  \right), & \mbox{for $d$ even}, \\
    O\left(n^{1 + \frac{(d-3)(d-1)}{2(d(d-1)-1)}}\log^{O(1)}n  \right), & \mbox{for $d$ odd},
\end{cases}
$$
\end{theorem}

\begin{figure}[htb]
    \begin{center}
        \scalebox{0.65}{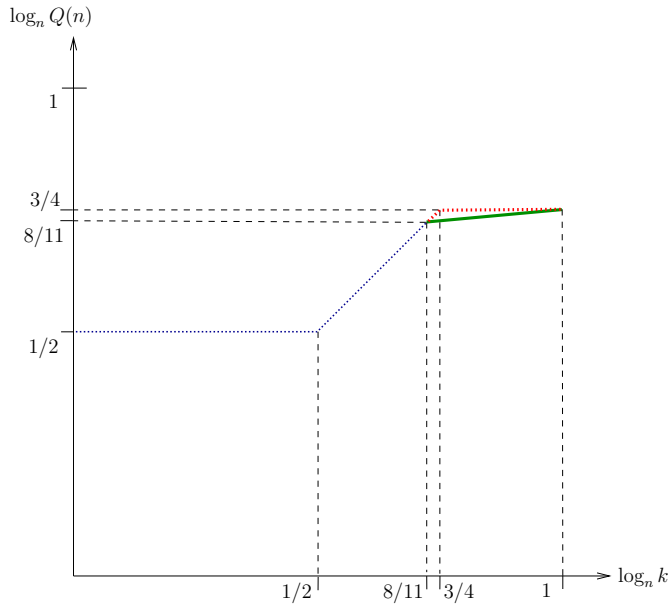}
        \caption{Graphical illustration of the output-sensitive improvement for $d = 4$.}
        \label{fig_output_sensitive1}
    \end{center}
\end{figure}

\noindent{\bf Remarks: (i)} To get some feeling for these bounds, we note that for $d = 3$ the query time is $O(\log n + w_{H})$ for $w_{H} \leq c_{1} n^{3/5}$ and $O(n^{1/2} w_{H}^{1/6} \log^{O(1)}n)$ otherwise; ignoring the logarithmic factors, both bounds balance out for $w_{H} = \Theta(n^{3/5})$. For $d = 4$, the query time is $O(n^{1/2}\log^{O(1)}n + w_{H})$ for $w_{H} \leq c_{1} n^{8/11}$ and $O(n^{2/3} w_{H}^{1/12} \log^{O(1)}n)$ otherwise. These bounds should be compared with the respective standard bounds $O(n^{2/3} \log^{O(1)}n)$ and $O(n^{3/4} \log^{O(1)}n)$ for $d = 3$ and $d = 4$ of \cite{mat92a}; our bounds coincide with these ``insensitive'' bounds when $w_{H} = \Theta(n)$. A graphical illustration that shows this improvement for $d = 4$ is given in Figure \ref{fig_output_sensitive1}. \\
\noindent{\bf (ii)} If $w_{H} > n/2$, denoting $\ol{H}$ as the complementary halfspace of $H$, we obtain the count by counting $n - |P \cap \ol{H}|$ (see also a preceding footnote). In this case the upper bound on the query time is obtained by replacing $w_{H}$ with $n - w_{H}$ in the above bounds. \\
\noindent{\bf (iii)} Clearly, a weak feature of the theorem is that the bound on the preprocessing cost is not near-linear. It approaches $O(n^{3/2})$ when $d$ is very large. For small values of $d$, it is near linear for $d = 3$ and is roughly $O(n^{14/11})$ for $d = 4$.

\noindent{\bf Proof.} We first describe the data structure, then the procedure for answering range counting queries and its analysis, then we analyze the storage used by the structure, and finally present and analyze the (somewhat involved) construction of the data structure.


\noindent{\bf Data structure.} We first construct an auxiliary data structure, based on Matou\v{s}ek's range reporting mechanism \cite{mat92b}, which allows us to report the points of $P$ in a query halfspace $H$ in time $O(n^{1-1/ \lfloor d/2 \rfloor} \log^{O(1)}n + w_{H})$, where $w_{H} = |P \cap H|$. This structure uses $O(n \log \log n)$ storage and can be constructed in $O(n \log n)$ time.

We next partition $P$ into a logarithmic number of subsets $P_{1}, \ldots, P_{l}$. For each $i$, $P_{i}$ consists of points which are $(c_{i}' k_{i})$-shallow in $P_{i-1}'$, where $P_{0}' = P$, $P_{i}' = P \setminus (P_{1} \cup \cdots \cup P_{i})$, for $1 \leq i \leq l-1$, and the values of the constant parameters $c_{1}', \ldots, c_{l}'$ will be set later. We set $k_{i+1} = 2k_{i}$, for $i \geq 1$, and set
\begin{equation} \label{eq_range_counting_k1}
k_{1} = c_{1} n^{1 - \frac{d-1}{d(d-1)-1}},
\end{equation}
for an appropriate constant $c_{1}$. With each $P_{i}$ we associate an auxiliary halfspace emptiness data structure, due to Matou\v{s}ek \cite{mat92b}. Putting $m_{i} = |P_{i}|$, the structure uses linear storage, requires $O(m_{i}^{1+\delta})$ preprocessing time, where $\delta > 0$ is arbitrarily small but fixed, and can test whether a query halfspace contains any point of $P_{i}$, in time $O(m_{i}^{1-1/ \lfloor d/2 \rfloor} 2^{O(\log^{*}m_{i})})$.

Next, as in the spanning tree construction (Theorem \ref{thm_spanning_T}), for each set $P_{i}$, we construct a simplicial partition $\Pi_{i}$, using Theorem \ref{thm_partition}, with $k = c_{i}'k_{i}$, in overall time $O(n \log n)$. With each class of each $\Pi_{i}$, consisting of a simplex $\Delta$ and a subset $P_{\Delta}$ of the corresponding $P_{i}$, we associate another auxiliary data structure, based on the standard simplex range counting technique of Matou\v{s}ek \cite{mat92a}. Putting $m_{\Delta} = |P_{\Delta}|$, this structure uses $O(m_{\Delta})$ storage and $O(m_{\Delta} \log m_{\Delta})$ preprocessing time, and can count the number of points of $P_{i}$ in a query halfspace in time $O(m_{\Delta}^{1-1/d}\log^{O(1)}m_{\Delta})$. This completes the description of our data structure.


\noindent{\bf Answering a counting query.} Given a query halfspace $H'$, our range counting algorithm proceeds as follows. We first run an emptiness test on each of the sets $P_{i}$, in ascending order of their indices, with respect to both $H'$ and its complementary halfspace $\ol{H'}$, using the auxiliary emptiness data structures associated with those sets. Assume that the test first returns {\bf true} at iteration $m$, for some $m \geq 1$. We then set $H$ to be the halfspace for which the emptiness test returned {\bf true}, and we refer to it as the query halfspace from now on. (If the original halfspace was the complement of $H$, we return the desired count by subtracting \footnote{If for some reason substraction is not allowed in the model of computation, we apply the querying procedure only to the input halfspace $H'$. We will then miss the opportunity to answer the query more efficiently when $n - w_{H}$ is small.} $|P \cap H|$ from $n$, as above.) Having a logarithmic number of sets, this step takes at most $O(n^{1-1/\lfloor d/2 \rfloor} \log^{O(1)}n)$ time.

Our analysis is based on Lemma \ref{lemma_missing_Pj}, given below, from which it follows that $H$ is disjoint from the convex hulls of all the sets $P_{j}$, for $j > U = \lceil \log (w_{H}/k_{1}) \rceil + 1$, and so it is guaranteed that $m \leq U + 1$. We distinguish between two cases. First suppose that $m \leq 2$. If $m = 1$, we have $P \cap H = \emptyset$, and thus we are done. If $m = 2$, we have $|P \cap H| \leq c_{1}' k_{1}$, and we use the reporting data structure associated with the full set $P$, to count the points in $P \cap H$ in time
$$
O\left(n^{1-1/ \lfloor d/2 \rfloor} \log^{O(1)}n + |P \cap H|\right),
$$
which establishes the first bound in (\ref{eq_range_counting_time}). Assume then that $m > 2$. It suffices to proceed only with the sets $P_{1}, \ldots, P_{m-1}$. For each such set $P_{i}$, with an associated simplicial partition $\Pi_{i}$, we check, for each class $(P_{\Delta}, \Delta) \in \Pi_{i}$, whether the simplex $\Delta$ is fully contained in $H$ or is crossed by its bounding hyperplane $h$. For a simplex $\Delta$ that is fully contained in $H$, we add $|P_{\Delta}|$ to the output count, whereas for $\Delta$ that is crossed by $h$, we use the auxiliary range counting data structure of \cite{mat92a}, associated with that node, to count $|P_{\Delta} \cap H|$. Repeating this procedure for $P_{1}, \ldots, P_{m-1}$, and adding up the resulting counts, we obtain the desired count $|P \cap H|$.

We now examine the value of $k_{1}$, as set in (\ref{eq_range_counting_k1}), and note that, for any halfspace $H$ with $w_{H} = \Theta(k_{1})$, the upper bound on the cost of answering the query with $H$ using the range reporting data structure is roughly the same, up to polylogarithmic factor, as the cost of answering the query using the range counting procedure at each of the simplices crossed by $h$. Indeed, the former cost is $O(n^{1-1/ \lfloor d/2 \rfloor} \log^{O(1)}n + k_{1})$, for $d \geq 4$, and the latter cost is
$$
O\left((n/k_{1})^{1-1/(d-1)} \cdot k_{1}^{1-1/d} \log^{O(1)}n \right) = O\left(n^{1-1/(d-1)} k_{1}^{1/d(d-1)} \log^{O(1)}n \right).
$$
This follows because $(i)$ for such values of $w_{h}$ we only need to query in $P_{1}$, $(ii)$ the number of simplices of $\Pi_{1}$ crossed by the plane $h$ bounding $H$ is $O((n/k_{1})^{1-1/(d-1)}\log^{O(1)}n)$ by Theorem \ref{thm_partition}, and $(iii)$ the cost of a query within each of the crossed classes is $O(k_{1}^{1-1/d}\log^{O(1)}k_{1})$. As is easily checked, our $k_{1}$ does indeed satisfy
$$
n^{1-1/\lfloor d/2 \rfloor} \log^{O(1)}n < k_{1} = \Theta\left(n^{1-1/(d-1)}k_{1}^{1/d(d-1)}\right),
$$
which makes the two bounds roughly the same, up to a polylogarithmic factor. For $d = 3$, the first bound is $O(\log n + k_{1})$, and the same relationship holds. For larger values of $w_{H}$, i.e., for $w_{H} > k_{1}$, recalling that $U = \lceil \log(w_{H}/k_{1}) \rceil + 1$, we use the second part of the data structure, recalling that, as above, $h$ crosses only $O((n/k_{i})^{1-1/(d-1)} \log^{O(1)}(n/k_{i}))$ simplices of $\Pi_{i}$ (see Theorem \ref{thm_partition}), to obtain an overall cost of
\begin{align*}
O\left(\sum_{i=1}^{U}(n/k_{i})^{1-1/(d-1)} k_{i}^{1-1/d} \log^{O(1)}n \right) & = O\left(\sum_{i=1}^{U}n^{1-1/(d-1)} k_{i}^{1/d(d-1)} \log^{O(1)}n \right) \\
& = O\left(n^{1-1/(d-1)} w_{H}^{1/d(d-1)} \log^{O(1)}n \right).
\end{align*}
This establishes the second bound on the query time, as given in (\ref{eq_range_counting_time}).

Comparing these bounds with the preceding techniques \cite{mat92a,mat92b}, we note that for $k \leq k_{1}$ we do not obtain any improvement --- we use in fact the same algorithm of \cite{mat92b}. However, for $k > k_{1}$, the bound on $Q(n)$ is smaller than the ``insensitive'' bound $O(n^{1-1/d} \log^{O(1)}n$ of \cite{mat92a}. Again, see Figure \ref{fig_output_sensitive1} for  graphical illustrations of our improvement. (Note that when $k = \Theta(n)$ our algorithm has the same asymptotic bound as Matou\v{s}ek's counting algorithm \cite{mat92a}.)

\noindent{\bf Storage.}
The auxiliary reporting data structure stored for the entire $P$ uses $O(n \log \log n)$ storage; see \cite{mat92b}. The emptiness data structures for each of the subsets $P_{i}$ are of size linear in the sizes of the respective subsets \cite{mat92a}, and so the total space that they consume is $O(n)$. The simplicial partitions constructed for the sets $P_{1}, \ldots, P_{l}$ consume a total of $O(n)$ storage, including the storage used by Matou\v{s}ek's standard range counting structure \cite{mat92a} within each simplex. Hence, the overall storage is $O(n\log\log n)$, as asserted.


\noindent{\bf Preprocessing time.} One of the main steps in the preprocessing (which is not required for the standard data structures of \cite{mat92a, mat92b}) is the construction of the subset of $k$-shallow points of a set $P$ of $n$ points in $\reals^{d}$, a step that we perform iteratively, to construct the sequence of sets $P_{1}, \ldots, P_{l}$. This step is involved (and costly) in higher dimensions, but we note that we only need to handle values of $k$ satisfying
$$
k > k_{1} = c_{1} n^{1 - (d-1)/(d(d-1)-1)},
$$
because for smaller values of $k$ we use instead the alternative range reporting machinery of \cite{mat92b}. For such values of $k$ we have
\begin{equation} \label{eq_range_counting_preproc}
\frac{n}{k} = O\left(n^{(d-1)/(d(d-1)-1)}\right),
\end{equation}
a property that will be crucial in guaranteeing the (relative) efficiency of the preprocessing.

So let $P$, $n$ and $k$ be fixed. The construction proceeds as follows. Put $r = n/k$ and choose a random sample $R$ of $c_{2}r\log r$ points of $P$, for an appropriate sufficiently large constant $c_{2}$. Let $K$ denote the interior of the intersection of all halfspaces that contain at least $|R| - c_{3}\log r$ points of $R$, for an appropriate constant $c_{3}$. By the centerpoint theorem (see \cite{mat02}), $K$ is nonempty for $c_{3}\log r < |R|/(d+1)$, which always hold for $c_{2}$ sufficiently large.

Before we describe and analyze the construction of $K$, we establish the following properties.

\begin{lemma} \label{lemma_shallow_net}
Let $P$, $R$, $K$ and $k$ be as above, and put $t = c_{3}\log(n/k) = c_{3}\log r$. Then an appropriate choice of the constants $c_{2}$ and $c_{3}$ guarantees that, with high probability, the following properties hold.
\begin{list} {\emph{(\roman{itemcounter})}}{\usecounter{itemcounter} \leftmargin=1em}
\item Every $k$-shallow hyperplane in $P$ is $t$-shallow in $R$.
\item Every hyperplane which is $(t + d)$-shallow in $R$ is $O(k)$-shallow in $P$.
\item Every point in $P \setminus K$ is $O(k)$-shallow in $P$.
\item Every point of $P$ which is $k$-shallow in $P$ lies in the complement $\ol{K}$ of $K$.
\end{list}
\end{lemma}
\noindent{\bf Proof.}
\noindent{({\romannumeral 1})} For $c_{2}$ sufficiently large, $R$ is a shallow $(1/r)$-net with high probability (see Section \ref{subsection_approximations} for the definition and properties of shallow nets). Since $r = 1/\eps = n/k$, the claim follows from Property (ii) of shallow $(1/r)$-nets, for an appropriate constant $c_{3}$.

\noindent{({\romannumeral 2})} This follows from Property (i) of shallow $(1/r)$-nets.

\noindent{({\romannumeral 3})} Let $p \in P \setminus K$, and let $H$ be a closed halfspace containing $p$, disjoint from $K$, and parallel to a closed halfspace $H_{0}$, which supports one of the open facets of $K$ ($H_{0}$ might coincide with $H$). By the definition of $K$, $H_{0}$ contains at least $|R| - t$ points of $R$. Since at most $d$ points lie on its bounding hyperplane, its  complementary closed halfspace $\ol{H_{0}}$ contains at most $t + d$ points of $R$, and thus, by Property (ii), it is $O(k)$-shallow in $P$, implying that $p$ is $O(k)$-shallow in $P$, as claimed.

\noindent{({\romannumeral 4})} We argue that no point $p \in P \cap K$ can be $k$-shallow in $P$. Indeed, let $p$ be such a point and let $H$ be a closed halfspace, whose bounding hyperplane $h$ contains $p$. By construction, since $h$ intersects $K$, it follows that $|H \cap R| > t$. By Property (i), we have $|H \cap P| > k$. Since this holds for every halfspace $H$ containing $p$, $p$ is not $k$-shallow in $P$, a contradiction which implies the claim. $\Box$

We next derive the following lemma, promised earlier, which is crucial for the efficiency of the query mechanism, and whose proof is based on the properties established in Lemma \ref{lemma_shallow_net}.

We define the {\em depth} of a point $p \in \reals^{d}$ as the minimum number of points of $P$ contained in a closed halfspace whose bounding hyperplane passes through $p$.

\begin{lemma} \label{lemma_missing_Pj}
Let $P$, its partition into $P_{1}, \ldots, P_{l}$, and the parameters $k_{1}, \ldots, k_{l}$ be as above. Let $H$ be a halfspace in $\reals^{d}$ of weight $w_{H}$. Then $H$ misses the convex hulls of the sets $P_{2}, \ldots, P_{l}$, if $w_{H} \leq k_{1}$, and of the sets $P_{j}$, for $j > U = \lceil \log (w_{H}/k_{1}) \rceil + 1$, otherwise.
\end{lemma}

\noindent{\bf Proof.}
First assume that $w_{H} \leq k_{1}$. Since the points in $P \setminus P_{1}$ are at least $(k_{1} + 1)$-deep, $H$ cannot contain any of them, and it therefore misses the convex hulls of the sets $P_{2}, \ldots ,P_{l}$, as claimed. We may thus assume that $w_{H} > k_{1}$, and so $U \geq 2$. Assume that $H$ crosses the convex hull of a set $P_{j}$, for $j > U$. Thus, necessarily, there exists a point $q \in P_{j}$ which is $w_{H}$-shallow. On the other hand, by Lemma \ref{lemma_shallow_net} ({\romannumeral 4}), $P_{j}$ consists of points which are at least $(k_{j-1}+1)$-deep in $P_{j-1}'$, and clearly so they are in $P$. Thus the depth of $q$ is at least
$$
k_{j-1}+1 > 2^{j-2}k_{1} \geq 2^{(\lceil \log (w_{H}/k_{1}) \rceil + 2) - 2} \cdot k_{1} \geq w_{H}.
$$
That is, $q$ is at least $(w_{H}+1)$-deep in $P$, in contradiction to its being $w_{H}$-shallow. $\Box$

Suppose that we have a procedure for constructing $K$. Then the construction of the sets $P_{i}$ proceeds as follows. Starting with the initial set $P$, put $\tilde{P} = P \setminus K$. Lemma \ref{lemma_shallow_net} (iii) implies that all the points of $\tilde{P}$ are $O(k)$-shallow in $P$ and thus also in $\tilde{P}$. We now iterate this process, as follows. At step $i$ we apply this construction to the subset $P_{i-1}'$ of the remaining points of $P$ (with $P_{0}' = P$), and take the resulting set $\tilde{P}$ as the next set $P_{i}$ in the sequence, setting $P_{i}' = P_{i-1}' \setminus P_{i}$; that is, $P_{i}'$ is the portion of $P_{i-1}'$ inside the corresponding convex region $K$. We stop the process when $P_{i}' = \emptyset$ or when $|P_{i}'|/k_{i}$ drops below a sufficiently large constant, according to the Partition Theorem (Theorem \ref{thm_partition}), which happens after $O(\log(n/k_{1}))$ steps. If $P_{i}' \neq \emptyset$ upon termination, we put $P_{i+1} = P_{i}'$.

We finally describe the construction of $K$, thus concluding the description of the new aspects of our data structure construction. We pass to the dual space, where each point $p \in P$ is mapped to a hyperplane $p^{*}$, using the standard duality which also preserves the above/below relationships (see, e.g., \cite{ed87}). Consider the arrangement $\mathcal{A} = \mathcal{A}(R^{*})$ of the hyperplanes dual to the points of the sample $R$. A $t$-shallow hyperplane $q$ in the primal space is mapped to a point $q^{*}$ which lies at level $\leq t$ in $\mathcal{A}$; more precisely, its level is either among the $t$ lowest levels of $\mathcal{A}$ or among its $t$ highest levels. We denote the former (resp., latter) collection of levels as $\mathcal{A}^{-}_{\leq t}(R^{*})$ (resp., $\mathcal{A}^{+}_{\leq t}(R^{*})$), but we use just $\mathcal{A}^{-}_{\leq t}$ (resp., $\mathcal{A}^{+}_{\leq t}$) to simplify the notation. As is well known \cite{mat91b} (and easy to verify), $K$ is mapped in the dual space to the (open) region $K^{*}$ enclosed between the upper convex hull of $\mathcal{A}^{-}_{\leq t}$, and the lower convex hull of $\mathcal{A}^{+}_{\leq t}$, in the sense that a point $v$ is in $K$ if and only if its dual hyperplane $v^{*}$ is contained in $K^{*}$.

We note that, by Clarkson and Shor \cite{cs89}, the combinatorial complexity of $\mathcal{A}^{-}_{\leq t}$ and $\mathcal{A}^{+}_{\leq t}$ is at most
\begin{align*}
u &= O\left(|R|^{\lfloor d/2 \rfloor} t^{\lceil d/2 \rceil} \right) = O\left(|R|^{\lfloor d/2 \rfloor} \log^{O(1)}n \right) \\
&= O\left((r\log r)^{\lfloor d/2 \rfloor} \log^{O(1)}n \right) = O\left((n/k)^{\lfloor d/2 \rfloor} \log^{O(1)}n \right).
\end{align*}
Using (\ref{eq_range_counting_preproc}) we have
$$
(n/k)^{\lfloor d/2 \rfloor} = O\left(n^{\frac{(d-1)\lfloor d/2 \rfloor}{d(d-1)-1}}\right),
$$
and the exponent in the right-hand side is smaller than $1/2$ when $d$ is odd, and only slightly larger than $1/2$ when $d$ is even, for $d \geq 3$. More precisely, $u$ satisfies
\begin{equation} \label{eq_range_counting_t_level}
u =
\begin{cases}
    O\left(n^{\frac{1}{2}\left(1 + \frac{1}{d(d-1)-1}\right)} \log^{O(1)}n \right), & \mbox{for $d$ even}, \\
    O\left(n^{\frac{1}{2}\left(1 - \frac{d-2}{d(d-1)-1}\right)} \log^{O(1)}n \right), & \mbox{for $d$ odd},
\end{cases}
\end{equation}
for any $d \geq 3$.

Agarwal et al. present in \cite{adbms98} a randomized algorithm for computing the $(\leq t)$-level of an arrangement of $n$ hyperplanes in $\reals^{d}$ within expected time $\Theta(u)$.

Let $V^{-}$ (resp., $V^{+}$) denote the set of vertices of $\mathcal{A}^{-}_{\leq t}$ (resp., $\mathcal{A}^{+}_{\leq t}$). We preprocess each of $V^{-}$, $V^{+}$ for halfspace emptiness queries, using the algorithm of \cite{mat92b}. Then, for each $p \in P$, we test whether $p \in K$ by testing whether the upper halfspace bounded by $p^{*}$ does not contain any point of $V^{-}$, and the complementary lower halfspace does not contain any point of $V^{+}$. All the points that fail the test are placed in $\tilde{P}$, and those that pass it are passed to the next iteration. In this way, we report the points in $\tilde{P}$ in expected time
$$
O\left(n \cdot u^{1-1/\lfloor d/2 \rfloor} \log^{O(1)}u \right).
$$
Recalling the bound (\ref{eq_range_counting_t_level}) for $u$, for $d$ even we report those points in
$$
O\left(n^{1 + \frac{(d-2)(d-1)}{2(d(d-1)-1)}} \log^{O(1)}n \right)
$$
expected time, and for $d$ odd in
$$
O\left(n^{1 + \frac{(d-3)(d-1)}{2(d(d-1)-1)}} \log^{O(1)}n \right)
$$
expected time. Having a logarithmic number of steps, we construct all the sets $P_{i}$ within the same asymptotic expected bound. We note that the overall time consumed by the construction of the other data structure components in each step is smaller than the bound above. We omit the description of these other steps of the construction, as they are identical to those given in \cite{mat92a, mat92b}. This completes the proof of Theorem \ref{thm_range_counting}. $\Box$


\end{document}